\documentclass{article}
\pdfoutput=1
\usepackage{graphicx}
\usepackage{natbib}
\usepackage{booktabs}
\usepackage{multirow}
\usepackage{bm}
\usepackage{amsmath}
\usepackage{amssymb}
\usepackage{geometry} 
\usepackage{color}

\newcommand\bu{\boldsymbol{u}}

\newcommand\bomega{\boldsymbol{\omega}}

\newsavebox{\astrutbox}
\sbox{\astrutbox}{\rule[-5pt]{0pt}{20pt}}

\newcommand\ba{\begin{eqnarray}}
\newcommand\ea{\end{eqnarray}}
\newcommand\be{\begin{equation}}
\newcommand\ee{\end{equation}}
\def\eqp#1{(\ref{eq:#1})}
\def\eql#1{\label{eq:#1}}

\newcommand\displayskip{\hskip 0.3truecm}

\newcommand\rmcore{\mathrm{core}}
\newcommand\rmDS{\mathrm{DS}}

\newcommand\rmp{\mathrm{p}}
\newcommand\rmlayer{\mathrm{layer}}
\newcommand\rmmax{\mathrm{max}}

\newcommand\rmmodel{\mathrm{model}}
\newcommand\rmslug{\mathrm{slug}}

\newcommand\sprime{s^\prime}

\newcommand\psistar{\psi^*}
\newcommand\tstar{t^*}
\newcommand\zetastar{\zeta^*}

\newcommand\psistarmax{\psistar_\rmmax}
\newcommand\zetastarmax{\zetastar_\rmmax}

\newcommand\Cmodel{C_\rmmodel}
\newcommand\deltacore{\delta_{\rmcore}}
\newcommand\deltalayer{\delta_\rmlayer}
\newcommand\p{\partial}
\newcommand\Rey{\mathrm{Re}}
\newcommand\threequart{\ensuremath{{\textstyle\frac{3}{4}}}}
\newcommand\sevenquart{\ensuremath{{\textstyle\frac{7}{4}}}}
\newcommand\psih{\widehat{\psi}}
\newcommand\psimax{\psi_\rmmax}
\newcommand\zetamax{\zeta_\rmmax}

\newcommand\Up{U_\mathrm{p}}
\newcommand\zetah{\widehat{\zeta}}
\newcommand\zetahf{\zetah_\mathrm{fit}}
\newcommand\zetahftwo{\zetah_\mathrm{fit2}}
\newcommand\zetahm{\zetah_\mathrm{model}}

\newcommand\bfx{\bm{x}}
\newcommand\bfxh{\widehat{\bfx}}

\newcommand\LDlim{(L/D)_\mathrm{lim}}

\newcommand\Et{\widetilde{E}}

\newcommand\CFM{C_\mathrm{FM}}
\def\order#1{{\cal O}\left(#1\right)}

\title{Advective balance in pipe-formed vortex rings\footnote{\textit{J. Fluid Mech.} (2018) \textbf{836}, 773--796.}}

\author{Karim Shariff\footnote{NASA Ames Research Center} and Paul S. Krueger\footnote{Southern Methodist University}}%

\begin{document}

\maketitle

\begin{abstract}
Vorticity distributions in axisymmetric vortex rings produced by a piston-pipe apparatus are numerically studied over a range of Reynolds numbers, $\Rey$, and stroke-to-diameter ratios, $L/D$.  It is found that a state of advective balance, such that $\zeta \equiv \omega_\phi/r \approx F(\psi, t)$, is achieved within the region (called the vortex ring bubble) enclosed by the dividing streamline.   Here $\zeta \equiv\omega_\phi/r$ is the ratio of azimuthal vorticity to cylindrical radius, and $\psi$ is the Stokes streamfunction in the frame of the ring.  Some but not all of the $\Rey$ dependence in the time evolution of $F(\psi, t)$ can be captured by introducing a scaled time $\tau = \nu t$, where $\nu$ is the kinematic viscosity.  When $\nu t/D^2 \gtrsim 0.02$, the shape of $F(\psi)$ is dominated by the linear-in-$\psi$ component, the coefficient of the quadratic term being an order of magnitude smaller.  An important feature is that as the dividing streamline ($\psi = 0$) is approached, $F(\psi)$ tends to a non-zero intercept which exhibits an extra $\Rey$ dependence.  This and other features are explained by a simple toy model consisting of the one-dimensional cylindrical diffusion equation.  The key ingredient in the model responsible for the extra $\Rey$ dependence is a Robin-type boundary condition, similar to Newton's law of cooling, that accounts for the edge layer at the dividing streamline. 
\end{abstract}


\section{Introduction}

\subsection{Scope and Motivation}\label{sec:Motivation}
 
The present work numerically studies the vorticity distribution in laminar vortex rings produced by a piston-pipe apparatus at Reynolds numbers high enough that an edge layer, thinner than the size of the ring, exists at the dividing streamline.  The diagnostics to be presented focus on the vorticity distribution in the region interior to the dividing streamline, called the vortex ring \textit{bubble}.  A model shows that the boundary condition provided by the edge layer is needed to account for an extra Reynolds number dependence, in addition to that accounted for by introducing a scaled time, $\tau = \nu t$.  However, the detailed structure of the edge layer and wake is not considered in the present work.  We hope that it is elucidated in the future.  For that effort, high Reynolds number falling drops \citep{Harper_and_Moore_1968} should provide a useful analogy: their structure also consists of an advectively balanced state in the interior surrounded by an edge layer that sheds a wake.

One motivation for studying the laminar vorticity distribution is that azimuthal instabilities are sensitive to its precise form \citep{Saffman_1978}.  Another motivation is that the non-dimensional energy, $\Et \equiv E / \left(I^{1/2} \Gamma^{3/2}\right)$, which determines the limiting stroke-to-diameter ratio according to \cite{Gharib_etal_1998}, depends on the peakiness of the vorticity profile.  (Here $E$ is the kinetic energy, $I$ is the impulse, and $\Gamma$ is the circulation; density is set to unity.)  For example, Hill's spherical vortex which has uniform $\zeta \equiv\omega_\phi/r$ gives $\Et = 0.16$ while the experimentally measured value for the limiting ring is a much larger $\Et \approx 0.33$. 
In addition, the dividing streamline of Hill's vortex has zero oblateness $(a - b)/b$ while the thickest experimental ring has an oblateness of  $0.21$ as inferred from photographs in \cite{Gharib_etal_1998}, $a$ and $b$ being the semi-major and minor axes, respectively.  The discrepancies in energy and oblateness are both consequences of the peakiness of $\zeta$ in actual rings.

The family of steady vortex rings \citep{Norbury_1973} with uniform $\zeta$, in which Hill's vortex is the thickest member, has been used to model vortex ring formation \citep{Mohseni_and_Gharib98, Shusser_and_Gharib00, Linden_and_Turner01}.  Since thinner rings have larger $\Et$, the result in these studies is that the limiting ring corresponds to a thinner member of the family than Hill's vortex.  If one were to use a family of rings having a more realistic peaked vorticity distribution, the result would presumably be that the limiting experimental ring corresponds better with the thickest member of that family.  

The physics of vortex sheet roll-up at an edge implies a peaked distribution of $\zeta$.  
\cite{Saffman_1978} and \cite{Pullin_1979} obtain the form of $\zeta$ for thin rings (and in the inner region of the core) by using Kirde's (\citeyear{Kirde_1962}) theory for two-dimensional vortex sheet roll-up at an edge.  This theory gives the vorticity in terms of the hypergeometric function.  For thicker rings, there is a calculation of their formation from axisymmetric vortex sheet roll-up \citep{Nitsche_and_Krasny_1994}, but no corresponding prediction of the vorticity distribution.  We hope that this lack is addressed soon.

\subsection{Present work}

The vorticity equation is $D\zeta/Dt = 0$ for inviscid swirl-free axisymmetric flow.  Hence if the flow is steady in some uniformly propagating frame, we have
\be
\zeta = F(\psi), \eql{zeta_equals}
\ee
where $\psi$ is the Stokes streamfunction in the ring frame.  It is natural to wonder if viscous vortex rings generated in the laboratory obey \eqp{zeta_equals} and if so, what form $F(\psi)$ takes.  Knowing this, one could solve for the main structure of the ring (i.e., apart from the wake and edge layer) by solving an elliptic free boundary problem \citep{Eydeland_and_Turkington_1988}.  It was this possibility that motivated the present work. 

The axisymmetric Navier-Stokes simulations performed here show that throughout the vortex ring bubble (except very close to the dividing streamline),  a state of $\zeta \approx F(\psi, t)$ is reached after a period of relaxation.  However,  even after moderately large times, $F(\psi, t)$ is not a universal function and depends on Reynolds number, $\Rey$, and stroke-to-diameter ratio $L/D$.  The best that can be said is that $F(\psi, t)$ tends to an approximately linear function with a non-zero intercept, whose dependencies will be studied in the sequel.

For interpreting results, it is helpful to keep in mind some consequences of the vorticity equation.  We adopt a reference frame moving with the ring at speed $U(t)$.  Because the frame is decelerating, a fictitious force term $-\rho\dot{U}\bfxh$, is present on the right-hand-side of the equation for momentum per unit volume.  However, for uniform density, $\rho$, this term can be absorbed into the pressure: $p \to p + \rho\dot{U} x$.  Thus, the vorticity equation remains unchanged in the moving frame:
\be
\frac{\p\bomega}{\p t} + \bu\cdot\nabla\bomega = \bomega\cdot\nabla\bu + \nu\nabla^2\bomega. 
\eql{vort_orig}
\ee
For axisymmetric swirl-free flow in cylindrical coordinates (Batchelor \citeyear{Batchelor_1967}, pg. 602), equation \eqp{vort_orig} becomes 
\be
   \frac{D\zeta}{Dt} \equiv \frac{\p\zeta}{\p t} + u_x \frac{\p\zeta}{\p x} + u_r \frac{\p\zeta}{\p r}  = \nu \left(\frac{\p^2\zeta}{\p x^2} + \frac{\p^2\zeta}{\p r^2} + \frac{3}{r}\frac{\p\zeta}{\p r}\right), \eql{vort}
\ee
where
\be
   u_x = \frac{1}{r}\frac{\p\psi}{\p r}, \displayskip u_r = -\frac{1}{r}\frac{\p\psi}{\p x},
\ee
defines the Stokes streamfunction $\psi$.  Note that for inviscid flow $\zeta \equiv \omega_\phi / r$ is conserved following fluid elements; the $r$ in the denominator accounts for vorticity stretching.
Suppose that at a given instant we have $\zeta = F(\psi, t)$ in a certain region of the flow.  The simulations will show that this is approximately the case within the vortex ring bubble.  Then by direct substitution into \eqp{vort} one sees that the advection terms sum to zero, which is called advective balance.  This does not imply that the subsequent evolution is viscous.  The diffusion operator in \eqp{vort} destroys advective balance when the ring is not thin; see \S\ref{sec:future}\ref{item:destruction}.  It is suggested that maintenance of the condition $\zeta \approx F(\psi, t)$ requires that non-linear terms still be active.  A solution for the time evolution of $F(\psi, t)$ must take advective effects into account and \S\ref{sec:future}\ref{item:destruction} conjectures that this effect is shear dispersion which leads to averaging of the vorticity field on a closed streamline.  A similar suggestion was made by \cite{Rhines_and_Young_1983} for planar vortices.

\subsection{Previous efforts}

\cite{Berezovski_and_Kaplanski_1987} obtained a time-dependent \textit{Stokes-flow} vortex ring solution with a peaked vorticity distribution.  This solution has zero thickness at $\nu t = 0$ and tends to the self-similar solution obtained by \cite{Phillips_1956} as $\nu t \to \infty$.  No advection effects are present aside from a spatially uniform drift, which was obtained by \cite{Kaplanski_and_Rudi_2001} using the procedures of \cite{Kambe_and_Oshima_1975} and \cite{Rott_and_Cantwell_1993a, Rott_and_Cantwell_1993b}.  \cite{Kaplanski_and_Rudi_2005} obtain the value of $\nu t$ at which the non-dimensional energy $\Et$ of the  Stokes solution equals the value 0.33 of the experimental limiting ring.  Interestingly, at this value of $\nu t$, the non-dimensional circulation and ring speed also roughly match experimental values.  This suggests that the Stokes flow solution does capture some of the overall properties of high Reynolds number rings.

A virtue of the Stokes solution is that it fully includes curvature effects for vorticity diffusion.  However, some issues arise because vorticity advection is absent: (i) Since the vorticity field is not deformed by the curvature-induced strain, vorticity contours are not axially elongated as in actual high Reynolds number rings.  To overcome this difficulty, \cite{Kaplanski_etal_2012} introduce axial elongation and radial compression factors into the vorticity field of the Stokes solution.  The values of the two factors are selected by matching the non-dimensional energy and circulation of the model ring against values obtained from Navier-Stokes simulations.  (ii) An edge layer, where advection and viscous terms nearly balance, is missing.  (iii) Since viscosity enters the vorticity field of the Stokes solution only via the product $\nu t$, the solution cannot exhibit the extra Reynolds number dependence observed in the present results.  (iv) Since the viscous term acts to destroy advective balance, the Stokes ring cannot in general satisfy advective balance except at early times when the ring is thin.

Among other results, \cite{Fukumoto_and_Moffatt_2000} obtained (from the Navier-Stokes equations) an asymptotic solution for the vorticity and streamfunction of a diffusing vortex ring to second order in the slenderness ratio.  The solution was stated in terms of ordinary differential equations with respect to distance from the core.   At second order, both vorticity and streamfunction contours become elliptical.  In the future, it would be worthwhile to probe this solution from the point of view of advective balance.

Finally, we note the work of \cite{Couder_and_Basdevant_1986} who studied the generation of vortex pairs by two-dimensional von K\'arm\'an wakes.  They briefly analyzed the functional form of the vorticity in terms of the streamfunction by using a scatter plot, and found it to be linear, which corresponds to the Lamb dipole \citep[][p.~535]{Batchelor_1967}.

\section{Simulation set-up and post-processing}

The axisymmetric and incompressible Navier-Stokes equations were solved using the commercial Fluent package.  Advection terms were discretized using the quadratic upstream interpolation (QUICK) scheme of B.P. Leonard (\citeyear{Leonard79}) which obtains values at cell faces using a linear combination of central and upwind interpolation.  The iterative SIMPLE algorithm was selected to obtain, at the end of a time-step, a consistent pressure-velocity pair that simultaneously satisfies the space-time discrete momentum equation and discrete mass conservation; see the text by \cite{Ferziger_and_Peric_2002}. 
The implicit Euler scheme with linearization of the non-linear term was selected for time discretization.  Dissipative schemes require a grid refinement study to ensure reliability of the results and this is described in Appendix~\ref{sec:refine}.

\begin{figure}
  \centerline{\includegraphics[width=4truein]{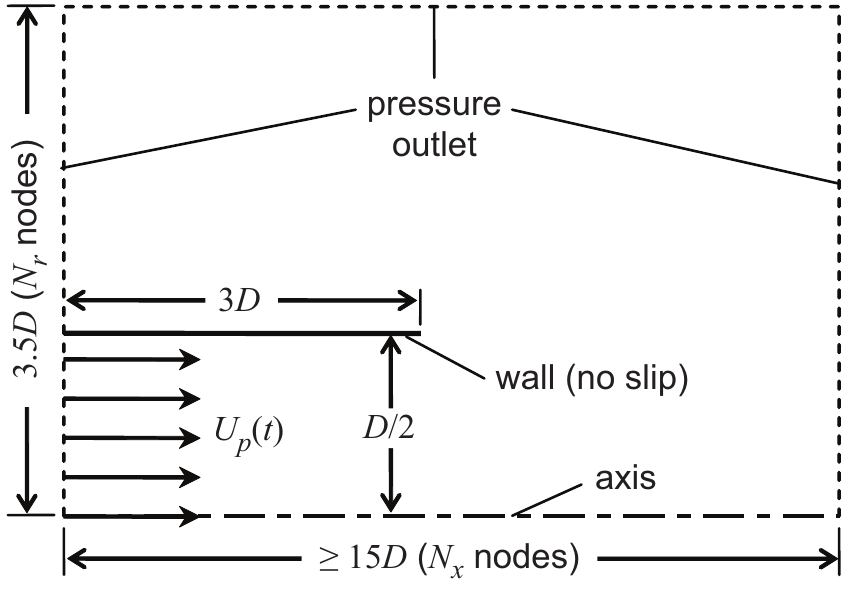}}
  \caption{Computational domain.  Not to scale.}
\label{fig:domain}
\end{figure}
Figure~\ref{fig:domain} shows the computational domain.  It consisted of a pipe of diameter $D$ and length $3D$.  The number of grid points is $N_x \times N_r = 1103 \times 251$.  The grid was non-uniform with the highest density near the tube wall and exit plane ($x = 0$).  The region, $r/D \le 1$ and $-0.5 \le x/D \le 8$, which we call the `vortex region', contained nearly all of the vorticity during the simulations and was chosen to have the highest resolution.  A no-slip condition was applied at the pipe wall and piston motion was simulated using a spatially uniform inlet velocity, $\Up(t)$, applied for $t \in [0, T]$.  The form of $\Up(t)$ was a trapezoidal pulse: the first and last 10\% of the pulse consisted of a linear acceleration and deceleration, respectively.  The maximum piston velocity is denoted $U_0$.  The resulting jet slug length was:
\be
   L = \int_0^T \Up(t)\, dt = 0.9 U_0 T. \eql{L}
\ee
A uniform pressure was specified at the outer boundaries.  The Reynolds number is defined as
\be
   \Rey \equiv \frac{U_0 D}{\nu}.
\ee

The Stokes streamfunction was determined by solving 
\be
	\frac{1}{r} \left(\frac{\partial^2 \psi}{\partial x^2}\right) + 
	\frac{\partial}{\partial r} \left(\frac{1}{r}\frac{\partial \psi}{\partial r}\right) = -\omega_\phi
	\eql{stream1}
\ee
in the frame of reference moving with the ring.  The solution for \eqp{stream1} was obtained in the neighborhood of the vortex ring by interpolating $\omega_\phi$ onto a regular grid and solving using a finite difference scheme with second-order truncation error.  To recast the velocity field (and hence the solution of \eqp{stream1}) in the frame of reference moving with the ring, the time varying ring velocity was determined from the derivative of the axial position of the ring centroid, defined as the centroid of vorticity greater than 60\% of the peak vorticity.  The time derivative was computed using a $4^{th}$-order finite-difference formula and smoothed using a Butterworth low-pass filter with a cutoff frequency of 10\% of the sample rate of the position data.

The procedure used to obtain $\zeta = F(\psi, t)$ was to consider 100 level curves of $\psi$ equi-spaced in the interval $0\le \psi \le \psimax$.  On each level curve, values of $\zeta$ at 100 points (equi-spaced in the angular direction centered at the location of the peak in $\zeta$) were obtained by spline interpolation from the solution grid.  Mean and standard deviations of $\zeta$ on each curve were then calculated.  This procedure, unlike the method of making a scatter plot of $\zeta$ versus $\psi$ previously used in the literature, facilitates further analysis.
\section{Overview and orientation}
\subsection{Overview}

Simulation cases were:
\begin{align*}
&\Rey = 1000\mathrm{\ and\ } L/D \in \left\{1, 4\right\};\\
&\Rey = 2000\mathrm{\ and\ } L/D \in \left\{0.5, 1, 2, 3, 4, 5, 6\right\};\\
&\Rey = 4000\mathrm{\ and\ } L/D \in \left\{1\right\}.
\end{align*}
Quantities normalized by piston speed $U_0$ and pipe diameter $D$ are denoted by an asterisk:
\be
   \tstar \equiv t U_0/D, \displayskip \zetastar \equiv \zeta / (U_0/D^2), \displayskip \psistar = \psi/U_0 D^2,
\ee
and a hat is used for $\zeta$ and $\psi$ normalized by their peak values (which decay in time):
\be
   \zetah \equiv \zeta/\zetamax(t), \displayskip \psih \equiv 1 - \psi/\psimax(t). \eql{hat_variables}
\ee
To diagnose the shape of $\zetah(\psih)$ at different instants, coefficients of the least-squares cubic
\be
   \zetahf(\psih, t) = 1 - a_1(t) \psih + a_2(t) \psih^2 + a_3(t) \psih^3, \label{eq:cubic}
\ee
fit to the data are obtained.  Only the mean value of $\zetah$ on each streamline was used for the fit.  To avoid the edge layer, the last value included in the fit is $\psih = 0.99$.
To ensure that $\zeta = F(\psi, t)$ to a good approximation, results of the fit and any quantity derived from it is plotted only if the standard deviation of $\zetah$ on every sampled streamline is $< 0.04$ for $0 \le \psih \leq 0.99$, and the rms error of the cubic fit is $< 0.005$.
After a referee's comment, a least squares fit was also performed to the form
\be
   \zetahftwo(\psih, t) = \exp\left(- b_1(t) \psih + b_2(t) \psih^2 + b_3(t) \psih^3\right), \label{eq:exp_cubic}
\ee
In only a small fraction of the fields (0 to 16\% for any given case) was it found to give a smaller rms error than the cubic.

A diagnostic of interest will be the value of $\zetah$ as the dividing streamline is approached from the interior of the ring bubble: $\zetah(\psih \to 1^-)$.  The use of a limit acknowledges the presence of an edge layer governed by boundary layer type behaviour which should match the interior flow.  The limit should be thought of in the same way that one thinks of the limit of potential flow past a body as the surface is approached, which should match the outer limit of the boundary layer.   The limit is obtained from the computations as
\be
   \zetah(\psih \to 1^-) = \zetahf(\psih = 1), \eql{operational}
\ee
which represents a slight extrapolation since the last value included in the fit is $\psih = 0.99$.

\subsection{Orientation}

\begin{figure}
\vspace{0.5truecm}
\centerline{\includegraphics[width=5.00truein]{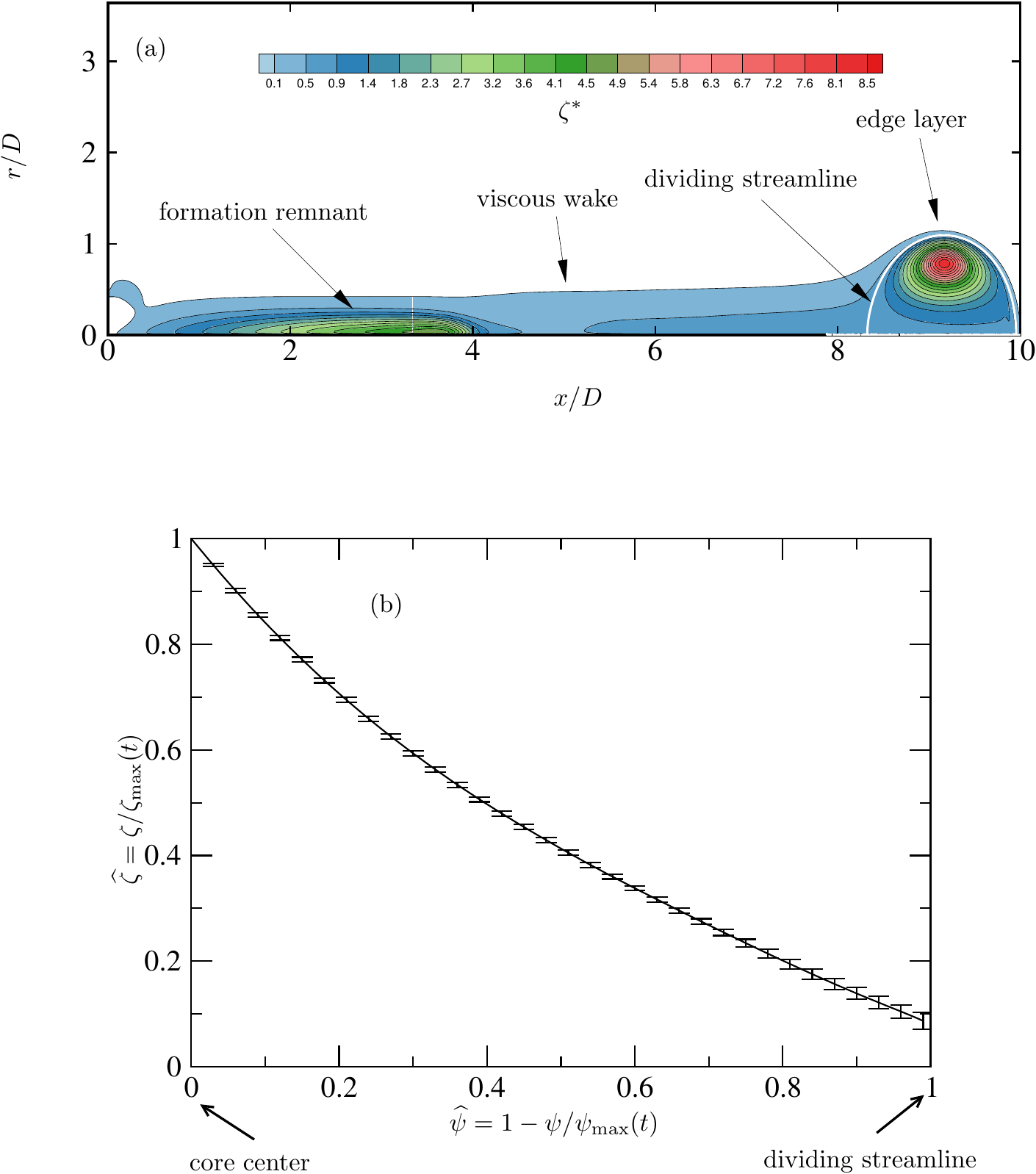}}
\vspace{0.70truecm}
\caption{(a) Contours of $\zetastar$ at $\tstar = 19.6$ ($\Rey = 2000, L/D = 4$).  White line: $\psi = 0$ contour in a frame moving with the ring bubble.
                 (b) The corresponding $\zetah(\psih)$ function in the region enclosed by the dividing streamline.  The width of the error bars is the standard deviation of $\zetah$ on each streamline and indicates the degree of advective balance on that streamline.}
\label{fig:zeta_contours}
\end{figure}
To orient the reader to typical flow behavior, Figure~\ref{fig:zeta_contours}a shows contours of $\zetastar$ for one case at a single instant.  As is well known \citep{Gharib_etal_1998}, for stroke ratios $L/D$ below a limiting value $\LDlim$, a single ring forms without leaving a trailing remnant.  For $L/D > \LDlim$, the ring cannot grow fast enough to accommodate all of the vorticity fed into it and the leading ring pinches-off from the feeding layer,  leaving behind a remnant trailing jet.  In the present simulations $\LDlim$ is between 3 and 4.  The stroke ratio, $L/D = 4$, for the case shown in Figure~\ref{fig:zeta_contours} is a little larger than $\LDlim$ and so there is a trailing remnant.  The white line is the dividing streamline, the contour of $\psi = 0$ in the frame moving with the ring bubble.  It separates fluid that is instantaneously moving with the ring from that flowing past.  At the dividing streamline, there exists a viscous edge layer where outwardly diffusing vorticity is partly swept into the wake, and partly re-enters the vortex ring bubble.  The viscous edge layer also includes the region within the bubble next to the symmetry axis.  It is on the symmetry axis that total circulation of the upper half plane is lost by vorticity diffusion.

Figure~\ref{fig:zeta_contours}b shows the corresponding function $\zetah(\psih)$ inside the ring bubble.  The width of the error bars is the standard deviation of $\zetah$ on each streamline; smaller values imply
better advective balance.  In this and future plots, the last plotted point is at $\psih = 0.99$ in order to avoid the edge layer where $\zetah \neq F(\psih)$.

\section{Model: One-dimensional radial diffusion with an edge layer}\label{sec:model}

A one-dimensional toy model based on vorticity diffusion in a cylindrical (and non-curved) vortex tube is presented here.  It was developed to explain the extra Reynolds number dependence observed for the vorticity on the dividing streamsurface, specifically, for the ratio $\zeta(\psi \to 0)/\zetamax(t)$.  The key ingredient in the model is a condition at the boundary of the vortex tube which accounts for the viscous layer at the dividing streamsurface of actual vortex rings.  The model assumes that the edge layer is circular which is not true even for thin rings.  Hence the model is not a mathematical solution in any asymptotic limit and should be viewed as being merely illustrative.

For a thin vortex ring, i.e., one whose core radius $\deltacore\ll R$ (the toroidal radius), variations in $r$ across the cross-section of the tube can be neglected, i.e., $r = R(1 + \order{\deltacore/R})$.  The vorticity dynamics is then locally (i.e., within the tube and in the co-moving frame) the same as for planar flow.  Therefore, if the vortex initially has concentric circular streamlines and vorticity contours, they will continue to remain so.  Therefore, it is reiterated that the model does not account for the effect of curvature-induced strain on the diffusion of vorticity in the interior of the vortex ring bubble.

A feature of circular streamline flow is that the advective terms are trivially zero and the azimuthal vorticity $\omega_\phi$ obeys the planar diffusion equation
\be
   \frac{\p\omega_\phi}{\p t} = \nu \frac{1}{s} \frac{\p}{\p s}\left(s\frac{\p\omega_\phi}{\p s}\right),  \displayskip 
   0 \le s \le s_\rmDS(t), \eql{diff}
\ee
where $s$ is radial distance measured from the center of the core in a meridional plane, and $s_\rmDS(t)$ is the distance to the boundary of the vortex tube; it is allowed to grow in time as the ring expands.  Equation \eqp{diff} is solved numerically.

Next comes an assumption that makes the treatment an illustrative model (even for a thin ring) rather than a rigorous analysis: \eqp{diff} will be applied all the way out to the dividing streamline (denoted using subscript DS) at which location, neither local two-dimensionality holds nor are streamlines circular.
The distance from the core center to the notional dividing streamline is denoted as $s_\rmDS(t)$ which represents roughly the average distance from the center of the ring to the actual dividing streamline.

At the dividing streamline of an actual ring there is an edge layer whose thickness $\deltalayer$ (averaged along the length of the streamline, say) is
\be
   \deltalayer\propto \left(\nu/\varepsilon\right)^{1/2}, \eql{deltalayer}
\ee
where
\be
   \varepsilon \propto \frac{\mathrm{ring\ speed}}{R(t)} \propto\Gamma(t) / 4\pi R(t)^2, \eql{strain}
\ee
is the characteristic strain-rate to which the layer is subject.  The boundary condition applied to the diffusion equation \eqp{diff} to model the edge layer is
\be
   \left.\frac{\p\omega_\phi}{\p s}\right|_{s = s_\rmDS} = -\Cmodel\frac{\omega_\phi(s = s_\rmDS)}{\deltalayer}, \eql{bc}
\ee
where $\Cmodel$ is a modeling constant which allows one to change all $\propto$ into $=$ signs.
Combining \eqp{deltalayer} and \eqp{strain} one sees that
\be
   \deltalayer = 2 \sqrt{\pi} (\Gamma(t)/\nu)^{-1/2} R(t) \eql{delta_layer2}
\ee
involves the instantaneous ring Reynolds number $\Gamma(t)/\nu$, a fact that will be important in the sequel.  

The asymptotic analysis of \cite{Fukumoto_and_Moffatt_2000} revealed that to third order in core thickness,  the radius of a diffusing ring grows linearly in time:
\be
   R(t) = R_0 + \CFM \nu (t - T) / R_0.
\ee
The rate of growth is different for the radius corresponding to the peak vorticity and the radius corresponding to the peak value of the co-moving streamfunction, the respective values of $\CFM$ being 4.59 and 2.59.
(Note that this difference implies loss of advective balance.)
Tests with the model revealed that the value of $\CFM$ influenced the decay rate of $\psistarmax(t)$ plotted in Figure ~\ref{fig:params_LD1}e.  We chose $\CFM = 2.9$ to obtain the best agreement between the model and simulation for $\psistarmax(t)$.
Ring expansion is not critical to the model and was included only to improve the postdiction of $\psistarmax(t)$.

We set $\Cmodel = 2.2$ and $s_\rmDS(t)/R(t) = 0.50$ based on a visual matching of model and simulation results for the $L/D = 1, \Rey = 1000$ case.  The boundary condition \eqp{bc} is of Robin-type because it involves a linear combination of value and derivative; it is analogous to Newton's law of cooling which is used to represent a layer of advective cooling at the surface of a heat conducting solid.  Ring vorticity is undergoing a similar advective-diffusive process in the edge layer.  Note that the boundary condition \eqp{bc} is non-linear in the vorticity due to the presence of $\Gamma(t)$, which is a functional of the vorticity.

The initial vorticity profile for the diffusion equation in terms of piston parameters is prescribed in Appendix \ref{sec:ic_model}.  This initial profile, which also depends on $\nu$, uses the theory of self-similar vortex sheet roll-up and is valid for thin cores.

To obtain the streamfunction as a diagnostic and to plot profiles of $\zeta(\psi)$, we first calculate $\zeta \equiv \omega_\phi/r$, which $\approx \omega_\phi/R$ to consistent order.  The circumferential velocity, $u_\theta(s)$, is then obtained from
\be
   u_\theta(s) = \frac{1}{s} \int_0^{s} \omega_\phi \sprime\, d\sprime.
\ee
The co-moving two-dimensional streamfunction, $\psi_{2D}$, is obtained by integrating
\be
   u_\theta = -\frac{\p\psi_{2D}}{\p s},
\ee
from which, finally, the Stokes streamfunction is $\psi = R\psi_{2D} + C_1$ to consistent order.  The integration constant $C_1$ is chosen to make $\psi = 0$ at the notional dividing streamline $s = s_\rmDS$.

\section{$\zeta(\psi,  t)$ for a thin Gaussian ring}
\label{sec:zeta_psi_Gaussian}

A referee requested comparison of vorticity profiles from the simulation against those for a thin Gaussian ring \citep{Saffman_1970}.  This solution has circular vorticity contours which is true in the simulations only for small $L/D$ at early times.  Nevertheless, because it is a simple analytical model for a diffusing ring, its $\zeta(\psi)$ profile provides a useful reference even outside its range of validity.  It should be noted that while the experiments of \cite{Cater_etal_2004} obtained Gaussian-like vorticity profiles, they were thinner in the radial direction than the axial direction due to the effect of strain.

The relations necessary for plotting $\zeta(\psi,  t)$ for the Gaussian ring are presented below.  Note that in making a comparison with simulations, the relations will be evaluated all the way to the dividing streamline which is outside their range of validity, $s \ll R$, even for thin rings (as before, $s$ is the distance from the center of the core in a meridional plane).  The ring radius ($R$) and circulation ($\Gamma$) needed to evaluate the relations are, as for the model described in the previous section, obtained from the expressions in Appendix \ref{sec:ic_model}.

When $\sigma(t)$ (the core radius) $\ll$ $R$, the dynamics are locally two-dimensional and the two-dimensional Oseen (Gaussian) vortex
is a valid local solution for $\omega_\phi$, and therefore
\be
   \zeta(s, t) \equiv \frac{\omega_\phi}{r} = \frac{\Gamma}{\pi \sigma^2 r} \exp(-s^2/\sigma^2)\left[1 + \mathcal{O}\left(s/R\right)\right], \hskip0.25 truecm \sigma = \sigma(t).
   \eql{zeta_Gaussian}
\ee
The speed of such a ring, needed later, was calculated by \cite{Saffman_1970}  as
\be
   U(t) = \frac{\Gamma}{4 \pi R}\left[\log\left(\frac{8R}{\sigma}\right) - 0.558 + \mathcal{O}\left(\frac{\sigma^2}{R^2}\log\frac{\sigma}{R}\right)\right].
\ee
Integrating the circumferential velocity of Oseen's vortex gives its streamfunction:
\be
   \psi_{2D}(s, t) = -\frac{\Gamma}{2\pi} \left[\log \frac{s}{\sigma} + \frac{1}{2}E_1(s^2/\sigma^2)\right],
\ee
where $E_1$ is the exponential integral and the integration constant was chosen to non-dimensionalize the argument of the logarithm.
The local Stokes streamfunction for the vortex ring is
\be
   \psi(s, t) = R\psi_{2D}\left[1 + \mathcal{O}(s/R)\right] + C_2, \hskip 0.5truecm s \ll R. \eql{psi_Gaussian}
\ee
The constant $C_2$ is not disposable and is obtained by using the asymptotic matching rule described by \cite{Moore_1980} in his calculation of the speed
of an elliptical core ring.  The  rule is that in the region $\sigma \ll s \ll R$, equation \eqp{psi_Gaussian} should match the streamfunction
for a zero thickness ring in the region $s \ll R$, which is given by Moore's equation (2.27):
\be
   \psi_0 = \frac{\Gamma R}{2 \pi} \left[\log\left(\frac{8R}{s}\right) - 2 + \mathcal{O}\left(\frac{s}{R}\right)\right] -
   \frac{1}{2} U(t) R^2.
\ee
Matching gives
\be
   C_2 = \frac{\Gamma R}{2\pi}\left[\log\frac{8R}{\sigma} - 2\right] - \frac{1}{2} U(t) R^2.
   \eql{constant}
\ee
Equations \eqp{zeta_Gaussian} and \eqp{psi_Gaussian} implicitly give $\zeta(\psi)$ for the Gaussian ring.  The
slope of this function can be obtained explicitly at the core center:
\be
\frac{d\zeta}{d\psi} = \frac{1}{R^2} \frac{d \omega_\phi}{ds} \left(\frac{d\psi_{2D}}{ds}\right)^{-1} 
= - \frac{1}{R^2}\frac{1}{u_\theta(s)}\frac{d \omega_\phi}{ds} = \frac{4}{R^2\sigma^2(t)}
\ee
at $s = 0$.

The behaviour of the core size is specified as
\be
   \sigma(t)^2 = \sigma_0^2 + 4\nu (t - T),
\ee
where $\sigma_0$ is the core size at the end of the piston stroke.  To obtain $\sigma_0$, we note that for the hypergeometric profile (eq. \ref{eq:utheta} in Appendix~\ref{sec:ic_model}) the radius of peak velocity is \citep{Saffman_1978} 
\be
   s_1 = 1.45 \left(4\nu T\right)^{1/2} \eql{s11}
\ee
at the end of the piston stroke.  On the other hand, the radius of peak velocity for the Gaussian vortex is
\be
   s_1 = 1.121\sigma_0. \eql{s12}
\ee
Equating \eqp{s11} and \eqp{s12} gives 
\be
\sigma_0 = 1.29 (4\nu T)^{1/2}. \eql{sigma_T}
\ee
Since the circulation of the Gaussian ring is assumed to be constant, growth of radius would imply increasing impulse $\propto \Gamma R^2$.  Hence its radius is kept fixed at its initial value.

\section{Results}

\subsection{$\Rey$ dependence at $L/D = 1$}
\begin{figure}
\vspace{0.75truecm}
\centerline{\includegraphics[width=6.0truein]{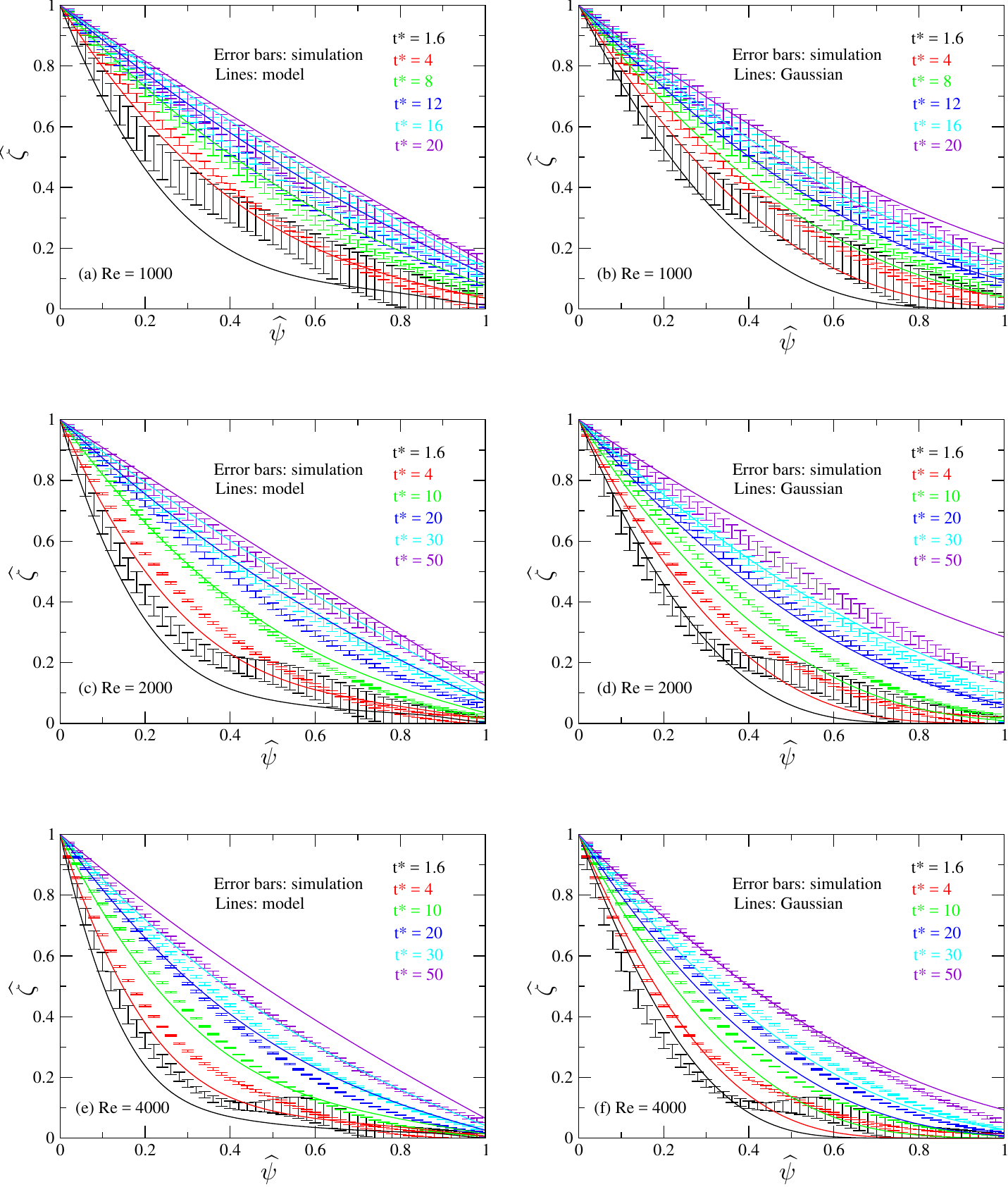}}
\caption{Profiles of $\zetah(\psih)$ for $L/D = 1$ and three different Reynolds numbers increasing downward.  The simulation results are shown using error bars; the width of the error bars equals the standard deviation of $\zetah$ on each streamline.  The left column of plots compares simulations with the model (lines).  The right column of plots compares simulations with the Gaussian ring.}
\label{fig:profiles_LD1}
\end{figure}

Figure~\ref{fig:profiles_LD1} shows profiles of $\zetah(\psih, t)$ for $L/D = 1$ and three Reynolds numbers.  The left column of plots compares simulation results (error bars) with the model (lines); the width of the error bars equals the standard deviation of $\zetah$ on each streamline.  
The right column of plots compares simulations with the Gaussian ring.
A small value for the standard deviation means that the vorticity is in approximate advective balance; this is achieved earlier and better for higher $\Rey$.  Vorticity diffusion causes the profiles to become less curved with time, which happens slower with increasing $\Rey$ as expected.  It is noteworthy that the value of $\zetah(\psih \to 1^-)$, i.e., as the dividing streamline is approached and ignoring the edge layer, is non-zero.  The value of $\zetah(\psih \to 1^-)$ increases with time and decreases with Reynolds number.
At early times, the profiles for the model and Gaussian ring do not have as much vorticity in the middle region of the bubble ($\psih \approx 0.5$) as the simulation does; this reflects inaccuracy in the initial condition provided by the vortex sheet roll-up theory (Appendix \ref{sec:ic_model}).
For the highest $\Rey$ case, a small secondary vorticity peak, which represents an outer spiral turn, is observed in this region.  The distance between adjacent spiral turns is larger than the diffusion length in this region \citep{Pullin_1979}.

The model initially lags the simulation in its development towards an approximately linear profile and then leads at later times.  Overall, the Gaussian ring gives a similar degree of agreement with the simulation as the model does, its most obvious error being a significant overshoot in the value of $\zetah(\psih \to 1^-)$ at late times.  This occurs because there is no edge layer to sweep away diffused vorticity.
\begin{figure}
\vspace{0.5truecm}
\centerline{\includegraphics[width=6truein]{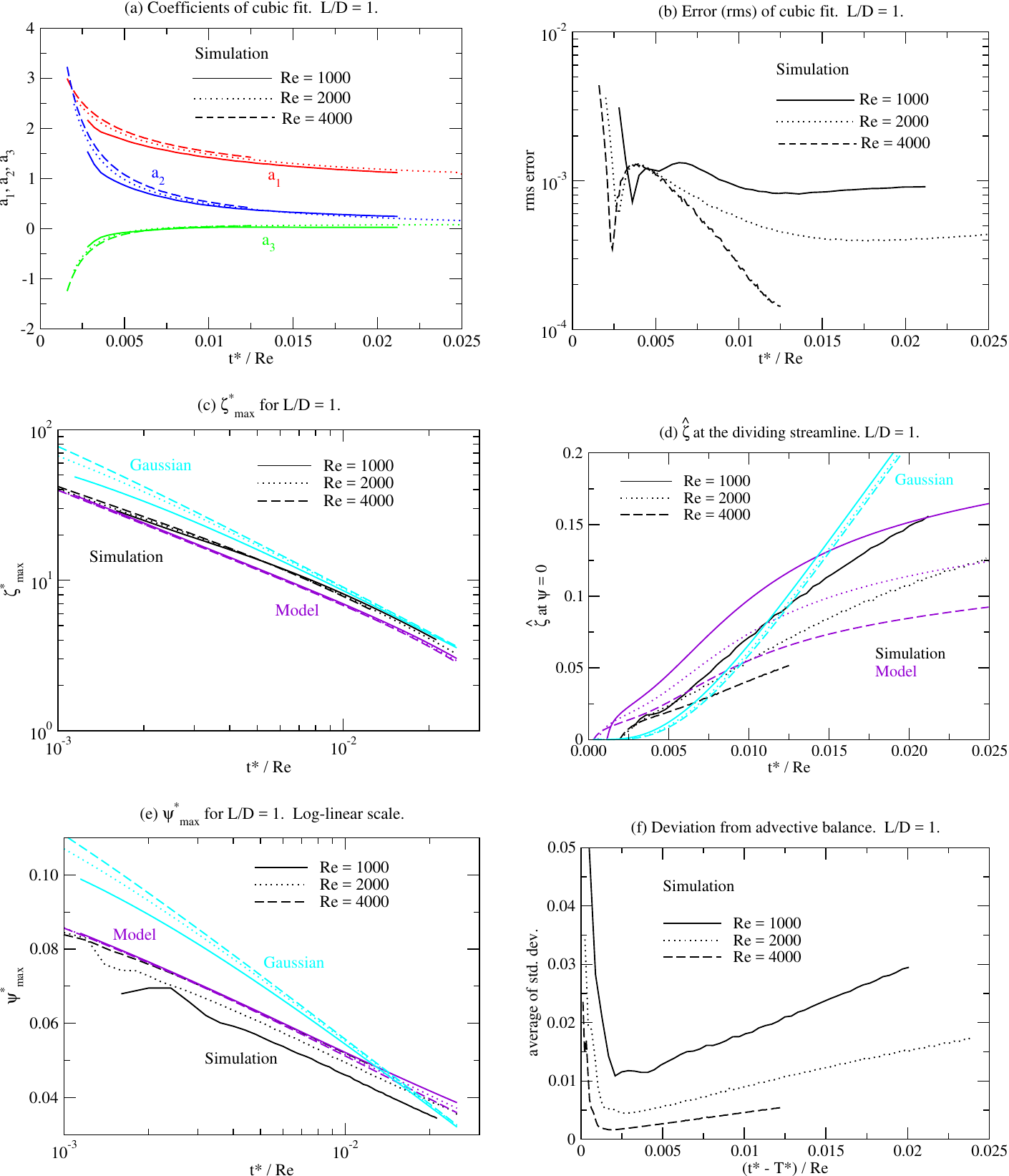}}
  \caption{Reynolds number dependence of profile parameters with time scaled by $\Rey$ ($L/D = 1$).}
\label{fig:params_LD1}
\end{figure}

Figure~\ref{fig:params_LD1} shows the evolution of various parameters of the vorticity profile with respect to $\tstar/\Rey = \nu t /D^2$.
Figure~\ref{fig:params_LD1}a displays the evolution of the cubic coefficients which are plotted only after the rms error (Figure \ref{fig:params_LD1}b) of the cubic fit becomes $ < 0.005$.
With time left unscaled, there was considerable difference in the curves (not shown) for the three $\Rey$ cases.  Scaling time by $\Rey$ reduces this difference due to the important role of viscosity in the overall evolution of $\zeta(\psi, t)$.  Some extra $\Rey$ dependence of the coefficients remains and this must ultimately be accounted for by Reynolds number effects in (i) the ring formation process, (ii) the boundary condition at the edge layer, and (iii) the competition between viscous destruction of advective balance and its restoration by non-linear terms.

As $\tstar/\Rey$ increases, the linear coefficient $a_1$ tends to approximately unity, while $a_2$ and $a_3$ tend to approximately zero.  This means that the profile becomes approximately linear with a small intercept equal to $(1 - a_1) + a_2 + a_3$ at the dividing streamsurface, $\psih \to 1^-$.

Figure \ref{fig:params_LD1}c shows that the decay of the peak vorticity ($\zetastarmax$) for the three $\Rey$ cases curves collapses quite well (but not perfectly) when time is scaled using $\Rey$.  The three violet colored curves (which are nearly coincident) show that the one-dimensional model tracks the simulation results quite well; the maximum error is about 10\%.

Figure~\ref{fig:params_LD1}d shows that the value of $\zetah$ as the dividing streamline is approached has a significant extra Reynolds number dependence.  It was the need to explain this that prompted development of the model with an edge-layer, whose results (violet lines) give the correct qualitative behaviour.

In the context of the model, the extra $\Rey$  dependence could arise from the presence of $\nu$ in both the initial condition (Appendix~\ref{sec:ic_model}) and boundary condition \eqp{bc}.  To ascertain which of the two was more important, control runs with the model were performed for the three Reynolds numbers in which the initial profile was fixed at the form given by the model for $\Rey = 2000$.
This resulted in only a very slight change compared to the model results in Figure~\ref{fig:params_LD1}d, indicating that most of the extra $\Rey$ dependence comes from the edge-layer boundary condition rather than changes in the initial profile.  The Gaussian ring (cyan lines) exhibits an insignificant amount of extra $\Rey$ dependence that is very different than observed in the simulations and the model.  This is because the Gaussian ring lacks an advective-diffusive edge layer and because $\nu$ enters the solution (mostly) via the product $\nu t$; the small extra $\Rey$ dependence is due to the presence of $4\nu T$ in the initial core size \eqp{sigma_T}.  The behaviour of $\zetah(t)$ for the Gaussian ring at the dividing streamline never goes through an inflection point and therefore has the opposite sign for the curvature at large times.

Figure \ref{fig:params_LD1}e shows the decay of the peak value, $\psistarmax$, of the streamfunction.  The best straight lines were obtained with a logarithmic abscissa.  The model agrees very well with the simulations for $\Rey = 4000$ while having a small constant error for the other two Reynolds numbers.  The small extra Reynolds dependence in $\psistarmax$ is not predicted by the model.

Figure~\ref{fig:params_LD1}f assesses the deviation from advective balance in the simulations using the standard deviation of $\zetah$ on each streamline averaged for $ 0 \le \psih \le 0.99$.  Note that the abscissa is measured from the end of the piston stroke at $t = T$.  At higher $\Rey$, advective balance is achieved to a greater degree and is destroyed more slowly with respect to $\tstar/\Rey$.
In Figure~\ref{fig:params_LD1}f  the rate of destruction of advective balance with respect to $\tstar/\Rey$ is roughly proportional to $\Rey^{-1}$, hence the rate with respect to $\tstar$ is roughly proportional to $\Rey^{-2}$.  The quadratic-in-$\nu$ rate of destruction suggests that the advective term plays a role in restoring advective balance.  One also observes that advective balance reaches the 0.02 threshold earlier (with respect to $\tstar/\Rey$) with increasing $\Rey$.  The increasing level of advective imbalance is likely a result of decreasing $\Gamma(t)/\nu$ with time.

\subsection{$L/D$ dependence at $\Rey = 2000$}
\begin{figure}
\vspace{0.75truecm}
\centerline{\includegraphics[width=6.0truein]{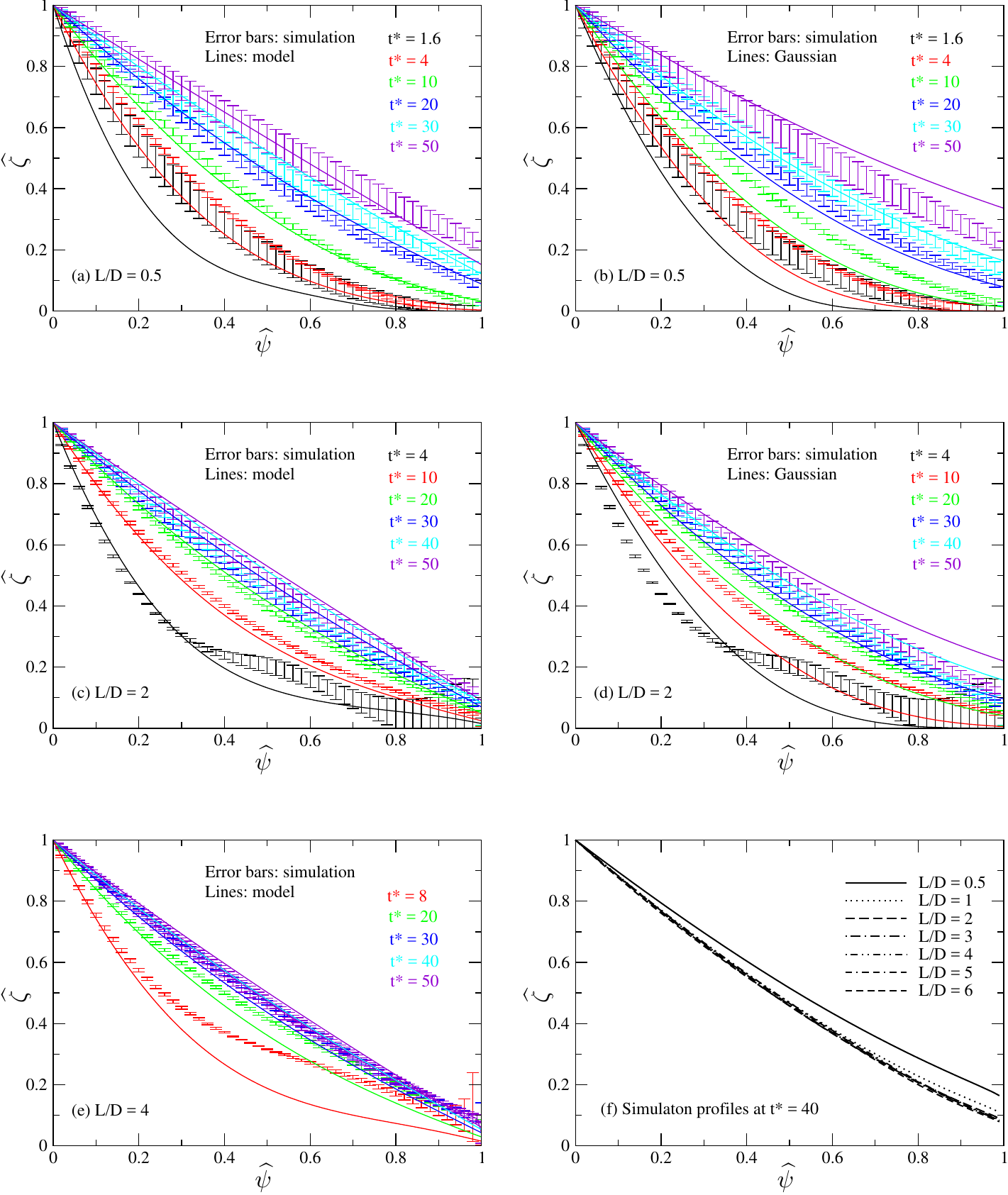}}
\caption{Profiles of $\zetah(\psih)$ showing the dependence on $L/D$ ($\Rey = 2000$).}
\label{fig:profiles_LD_dep}
\end{figure}

Recall that the limiting value of $L/D$ beyond which the ring trails a jet or sheds some vorticity during the formation process is $3 < \LDlim < 4$ in the simulations.  Model results are presented for only $L/D \leq 3$: according to current understanding,  there should be no $L/D$ dependence for $L/D \geq \LDlim$.  The simulations will indicate that this is not the case for $\psistarmax$.

Figure~\ref{fig:profiles_LD_dep} shows $\zetah(\psih, t)$ profiles.  The error bars indicate that for increasing $L/D$ (thicker cores) it takes longer for profiles to reach advective balance near the dividing streamline.  At late times, however, the smallest $L/D$ case ($L/D = 0.5$) is the least advectively balanced overall.  We shall see below that this is because it has the smallest ring Reynolds number $\Gamma/\nu$.
As observed in the Reynolds number study presented earlier (Figure~\ref{fig:profiles_LD1}), the development of model profiles initially lags and later leads the simulations.  Apart from the overshoot at late times, the Gaussian ring profiles (Figures~\ref{fig:profiles_LD_dep}b and d) agree with the simulations about as well as the model does.

Figure~\ref{fig:profiles_LD_dep}f compares simulation profiles for the different $L/D$ cases at a fixed late time of $\tstar = 40$.  The profiles for $L/D \ge 2$ are practically coincident.  The $L/D = 1$ (dotted) case is a little more diffuse, and the $L/D = 0.5$ (solid) case somewhat more diffuse than the rest.  This is because smaller $L/D$ rings have a smaller $\Gamma_0/\nu$ and are therefore more diffusive.
\begin{figure}
\vspace{0.5truecm}
\centerline{\includegraphics[width=6.2truein]{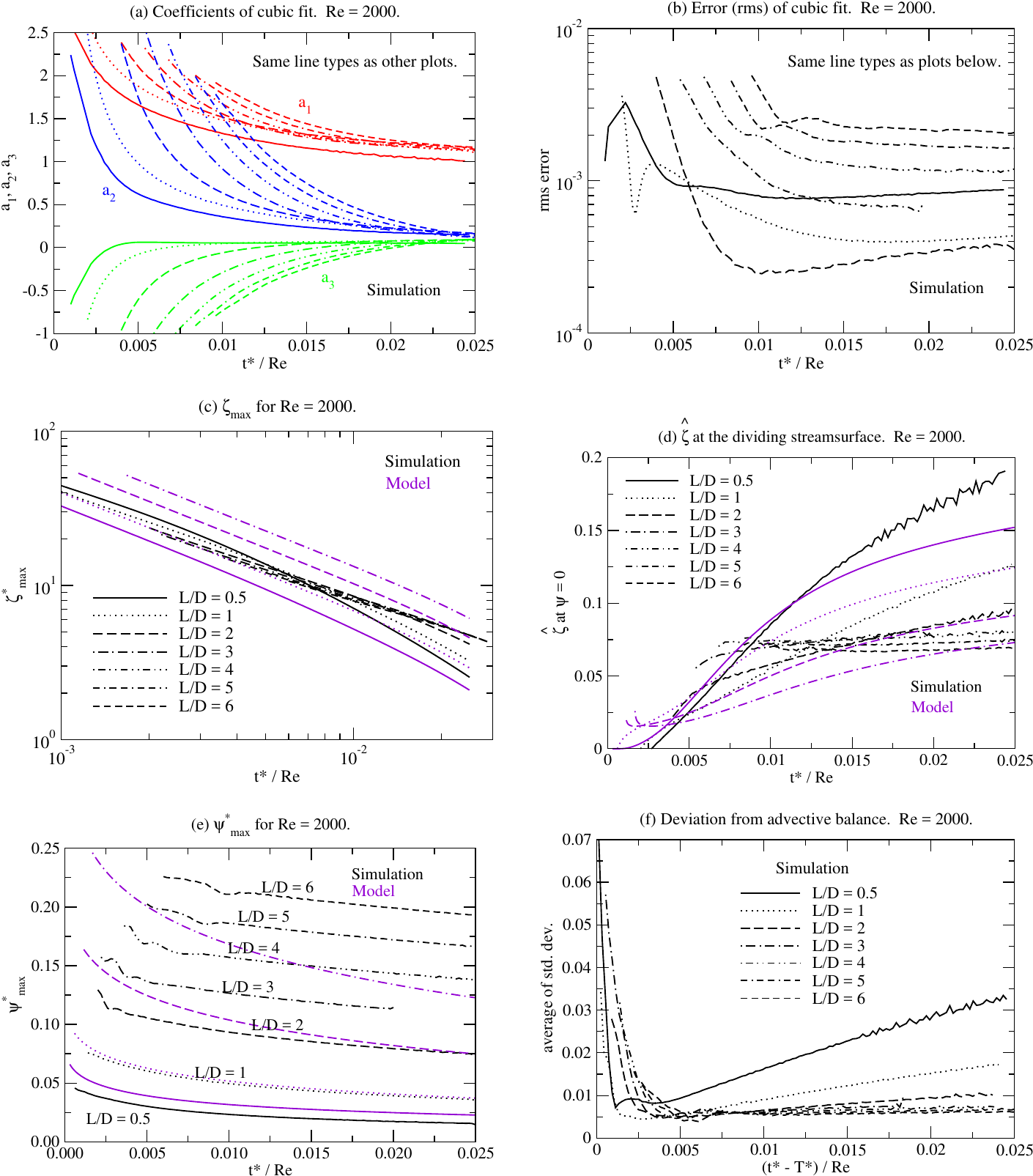}}
\caption{$L/D$ dependence of various parameters of the vorticity profile ($\Rey = 2000$).}
\label{fig:params_Re2000}
\end{figure}

The cubic coefficients in Figure~\ref{fig:params_Re2000}a show that differences in the shape of $\zetah(\psih)$ at early $\tstar/\Rey$ diminish later and the shape converges to a form that is weakly dependent on $L/D$; some variation and crossing of the curves at large times should be noted. 
The curve for $a_1$ and $L/D = 0.5$ (solid red line) has a noticeable shift relative to the other curves perhaps due to its low ring Reynolds number $\Gamma(t)/\nu$. 

The convergence noted in the previous paragraph is reminiscent of the study by \cite{Stanaway_etal_1988a} of viscous vortex rings which found (pp. 74--76 in their work) that, for different initial core thicknesses (but same $\Gamma_0, \nu$, and $R_0$), plots of normalized ring speed versus $\nu t / R_0^2$ all converged to approximately the same curve.  In their plots, $t$ is measured from a virtual origin when the ring had zero core thickness.

Figure~\ref{fig:params_Re2000}c displays the decay in peak vorticity, $\zetastarmax(t)$.  The model works best for $L/D = 1$ and has moderately large errors for the other $L/D$ cases.
These errors are present in the initial condition and this suggests improvement of the relations in Appendix \ref{sec:ic_model}.

Figure~\ref{fig:params_Re2000}d shows $\zetah$ at the dividing streamline.  At early times, a lower $L/D$ produces a smaller value of this quantity. This is expected since smaller $L/D$ leads to a thinner core.  However, a cross-over occurs and eventually smaller $L/D$ produces larger values of $\zetah(\psih \to 1^-)$ with unabated growth at the final instant of the simulation.  This is non-intuitive at first sight, and it is hypothesized that it is mainly due to smaller $L/D$ rings having a smaller initial ring Reynolds number $\Gamma_0/\nu$.  Since the trend of the dependence on $L/D$  at large times is reproduced by the model, we can use it to test the hypothesis.  We re-ran three $L/D$ cases ($L/D = 0.5, 1$, and 2) using the model, but set the initial circulation for all cases to the value for $L/D =2$.  Doing so reduced the range of variation in $\zetahm(\psih=1, \tstar/\Rey = 0.025)$ from $[0.091, 0.152]$ to $[0.091, 0.097]$.  Hence a substantial portion of the variation with respect to $L/D$ is due to changing $\Gamma_0/\nu$.

Next we hypothesize that this effect of $\Gamma_0/\nu$ (in the context of the model) enters mainly through the presence of $\Gamma(t)/\nu$ in the edge-layer boundary condition \eqp{delta_layer2}.  To verify this, a test was performed in which $\Gamma(t)$ in the model boundary condition was kept fixed in time at the $\Gamma_0$ value for $L/D=2$.  In this case the range of variation of $\zetahm(\psih=1, \tstar/\Rey = 0.025)$ for the three $L/D$ cases was very small, namely, $[0.076, 0.078]$.  We conclude from this that a substantial portion of the $L/D$ dependence of $\zetah(\psih \to 1^-)$ is due to the dependence of edge layer physics on $\Gamma(t)/\nu$, where $\Gamma_0$ depends on $L/D$.

To avoid clutter, results for the Gaussian ring have not been included in Figure~\ref{fig:params_Re2000}.  Suffice it to say, it does not provide better predictions than the model.

\begin{figure}
\vspace{0.5truecm}
\centerline{\includegraphics[width=3truein]{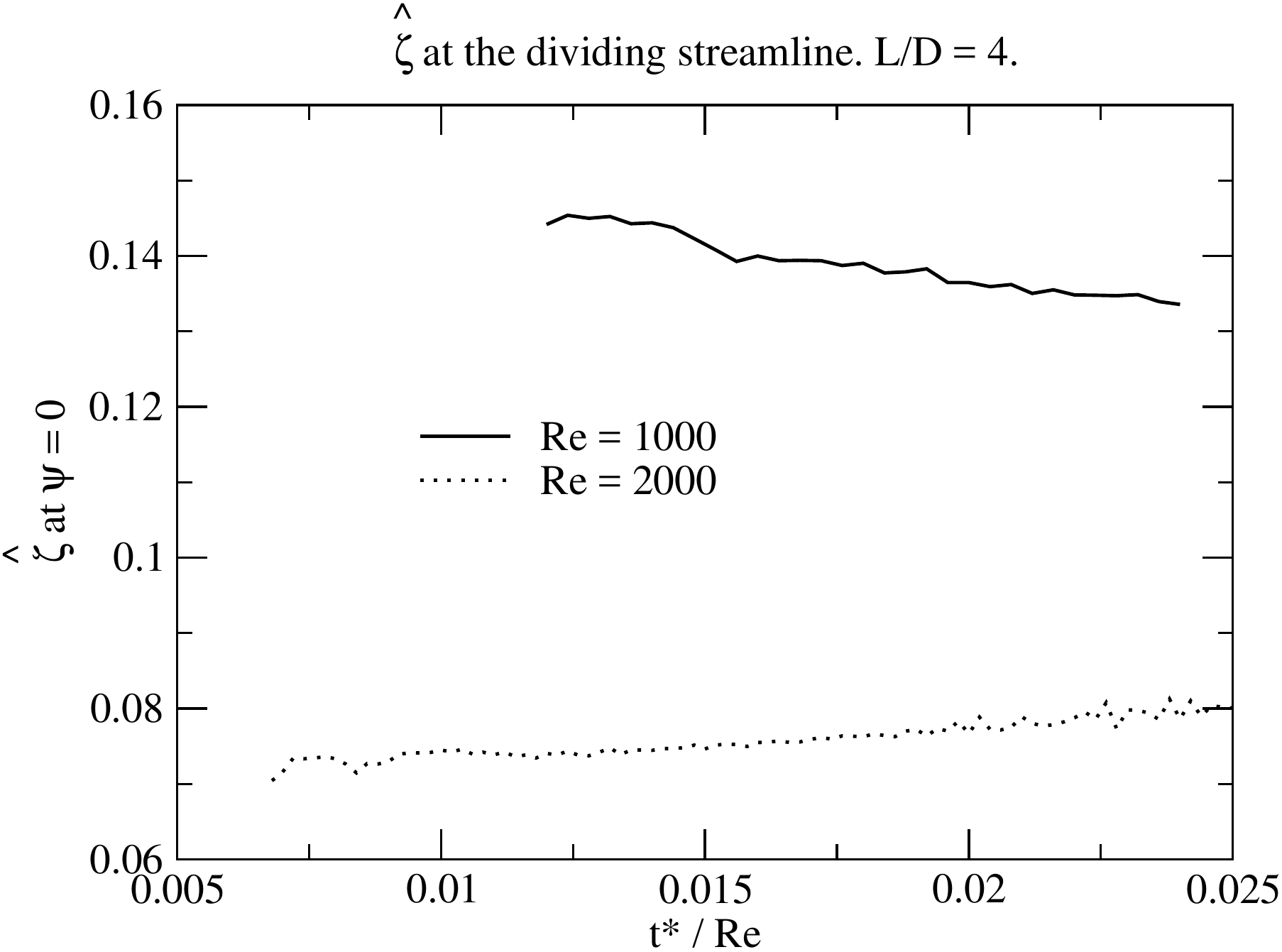}}
 \caption{Reynolds number dependence at $L/D = 4$ of $\zetah$ at the dividing streamline.}
\label{fig:zeta_hat_at_one_LD4}
\end{figure}
An important feature of the simulations which the model is unable to reproduce is that for $L/D \geq 3$, the temporal growth of $\zetah(\psih \to 1^-)$ saturates to a constant value.  This constant value has a weak dependence on $L/D$ for $L/D \geq 3$.  This is because $\Gamma_0/\nu$ loses its dependence on $L/D$ for $L/D > \LDlim$.  The flat value for $L/D = 4$ decreases with Reynolds number as shown in Figure~\ref{fig:zeta_hat_at_one_LD4}.  Although the model is inaccurate at this value of $L/D$, we hypothesize on the basis of previous model results that this dependence mainly reflects the change in $\Gamma_0/\nu$ which affects the edge layer.

Figure~\ref{fig:params_Re2000}e shows that the peak value of the streamfunction increases with $L/D$, a trend that is also captured by the model.  However, the value of $\psistarmax$ for $L/D = 3$ is over-predicted by a fair amount.

According to current thinking, the characteristics of the leading ring experimentally produced when $L/D\geq\LDlim$ should be independent of $L/D$; for example the circulation becomes independent of $L/D$ \citep{Gharib_etal_1998}.  However, Figure~\ref{fig:params_Re2000}e shows that $\psistarmax$ continues to increase with $L/D$.
Since $\psistar = 0$ at the dividing streamline, $\psistarmax$ represents the volumetric flux of recirculating flow in the bubble.  The mechanism that causes this quantity to increase
for $L/D > \LDlim$ should be investigated in the future.

Figure~\ref{fig:params_Re2000}f assesses the degree of advective balance as a function of $L/D$.  We observe that cases with smaller $L/D$ relax faster.  This is consistent with intuition: thinner rings have most of their vorticity in a region where local two-dimensionality holds and planar flows with circular streamlines remain in exact inviscid balance as they diffuse.  However, at later times, the smaller $L/D$ cases lose advective balance faster.  This is because smaller $L/D$ gives smaller $\Gamma_0/\nu$.  There is very little difference in the loss of advective balance for cases $L/D \ge 4$ because all these cases have approximately the same $\Gamma_0/\nu$.

The notion that the $L/D$ dependence is mostly due to changes in $\Gamma_0/\nu$ can also be approximately tested using the simulations in the following way.  Substituting the expressions for piston stroke \eqp{L} and slug circulation \eqp{Gamma_slug} into the empirical equation \eqp{Gamma0} for $\Gamma_0$ gives
\be
   \Gamma_0/\nu = .548 \Rey_L + 0.154 \Rey,  \displayskip L/D \le \LDlim, \eql{Gamma0_nu}
\ee
where $\Rey_L \equiv U_0 L/\nu$ is the Reynolds number based on piston stroke.  Equation \eqp{Gamma0_nu} shows that matching $\Rey_L$ approximately matches $\Gamma_0/\nu$.  In particular, the two cases $(\Rey = 2000, L/D = 0.5)$ and ($\Rey = 1000, L/D  = 1$) have the same $\Rey_L = 1000$ and \eqp{Gamma0_nu} gives their corresponding $\Gamma_0/\nu$ as 856 and 702, respectively.  
\begin{figure}
\includegraphics[width=2.7truein]{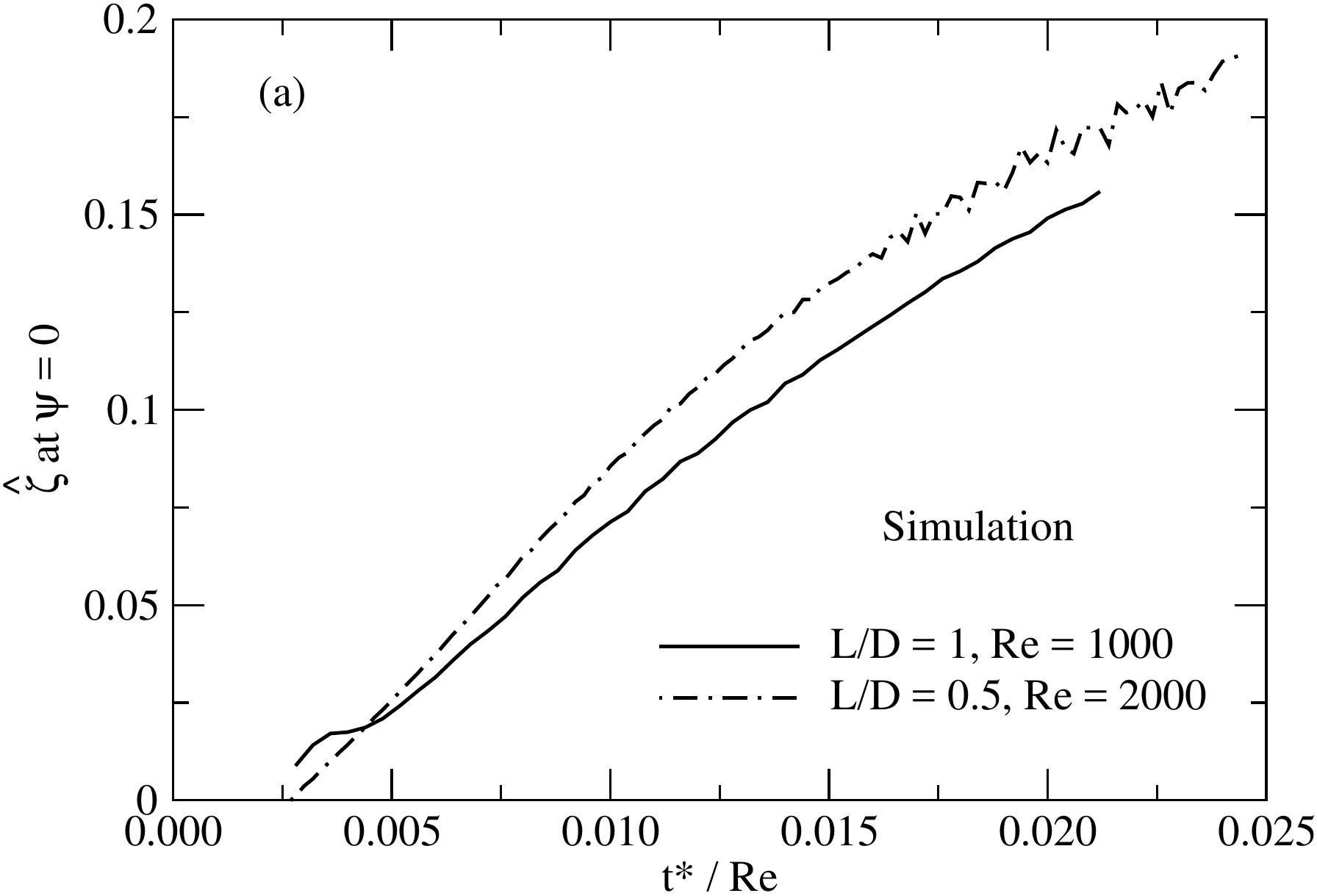}
\hfill
\includegraphics[width=2.7truein]{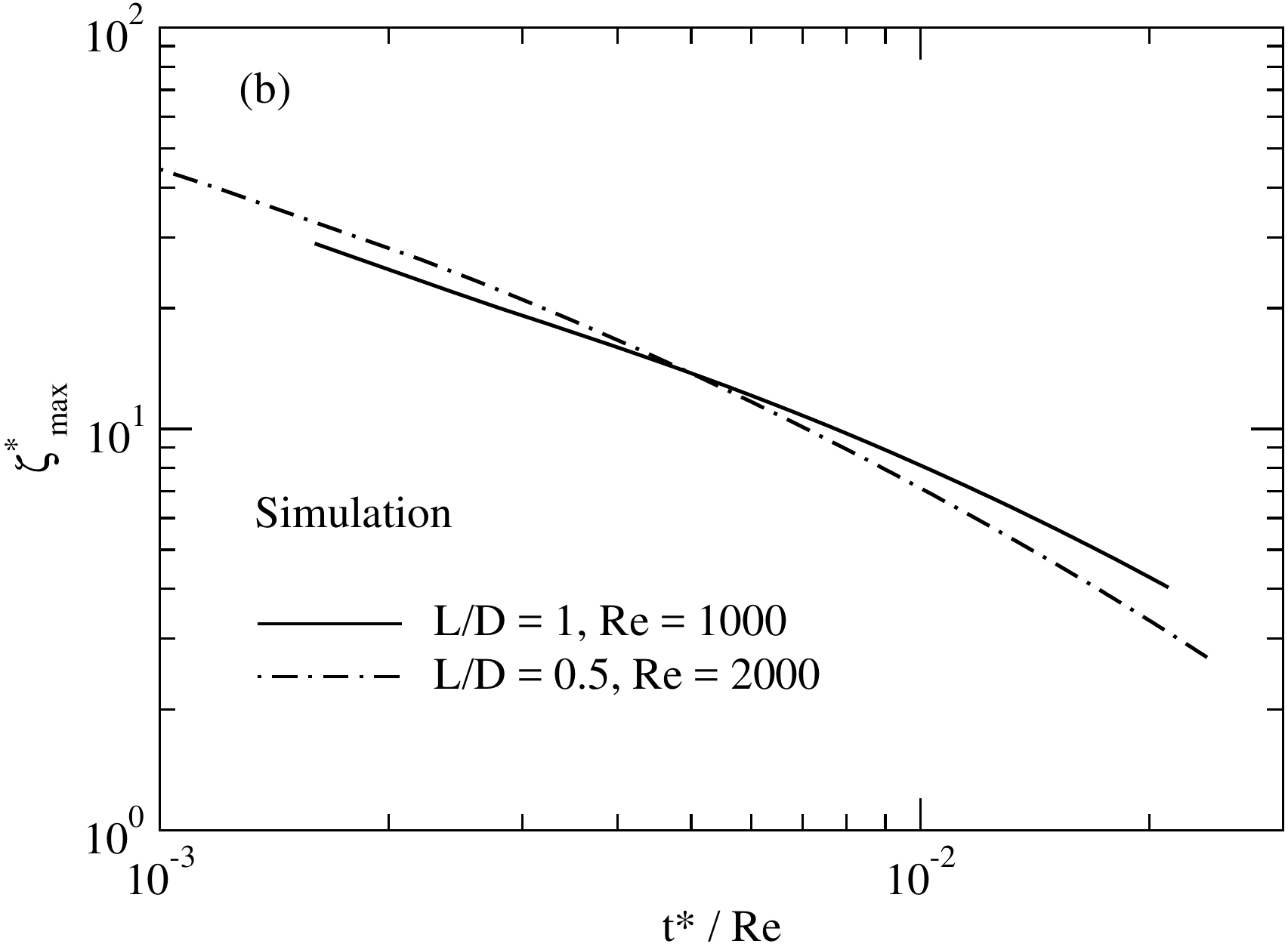}
\vskip 0.5truecm
\centerline{\includegraphics[width=2.7truein]{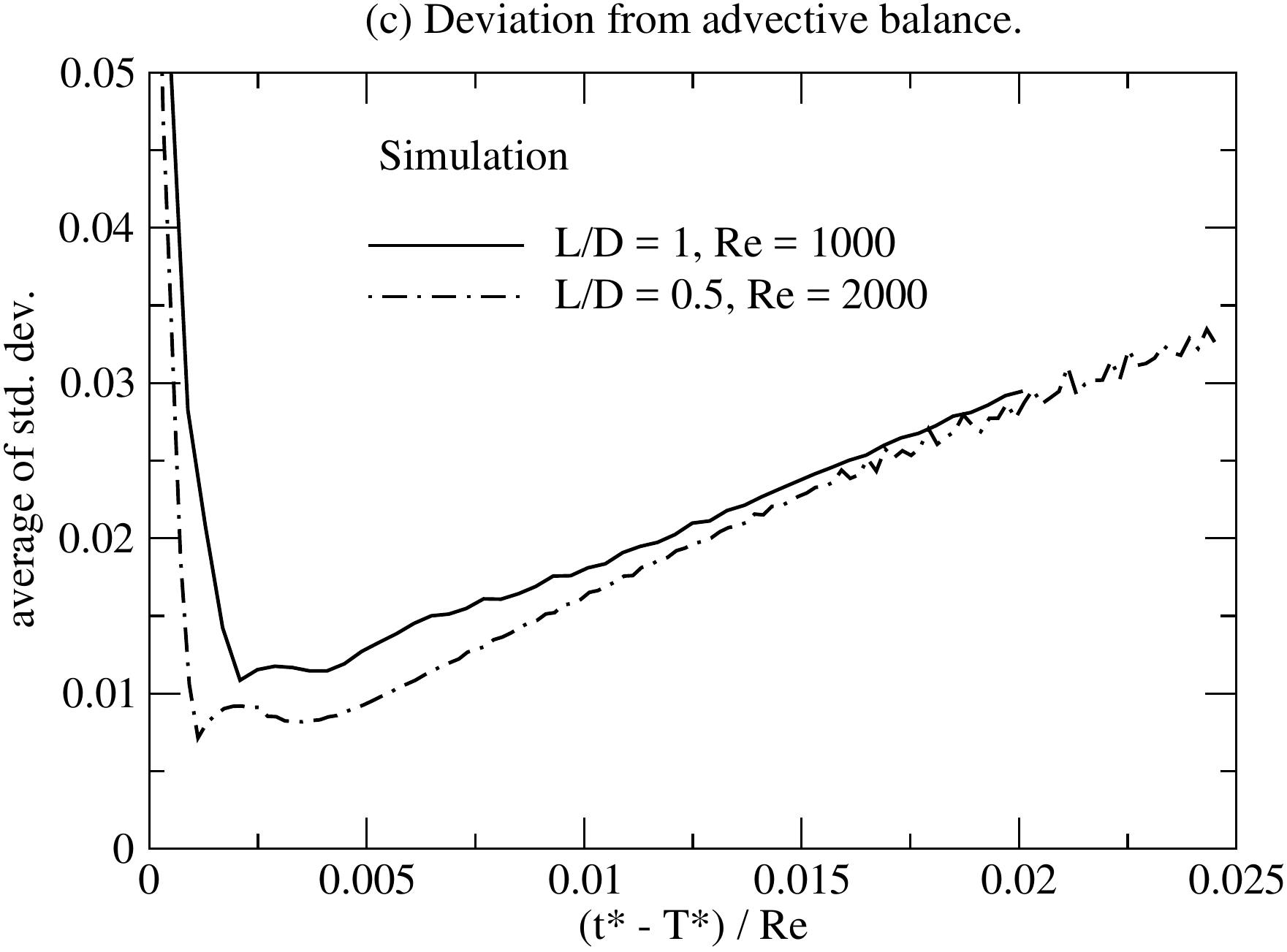}}
 \caption{Comparison of two cases that have similar $\Gamma_0/\nu$.}
\label{fig:hypothesis}
\end{figure}
Simulation results for the two cases are shown as the chain-dotted and solid black lines in Figure~\ref{fig:hypothesis}.  
Comparing them, one observes a reasonable correspondence in $\zetah(\psih \to 1^-)$, $\zetastarmax$,
and the rate of loss of advective balance.
The quantity $\psistarmax$ is not shown;  it cannot be expected to correspond in the two cases since it is affected by the core size to radius ratio.  

\section{Summary}\label{sec:Summary}

\begin{enumerate}

\item
After a period of relaxation, a state of approximate advective balance, $\zeta \approx \zeta(\psi, t)$, is achieved within the region bounded by the dividing streamline.

\item
The function $\zetah = \zetah(\psih, t)$ (where hats denotes quantities normalized by peak values) evolves in time from being convex to being approximately linear, but with a non-zero intercept as the dividing streamline is approached. 

\item
Introduction of the viscous time, $\tstar/\Rey = \nu t/D^2$,  captures some but not all of the Reynolds number dependence in the evolution of $\zetah = \zetah(\psih, t)$.
In particular, the value of $\zetah$ at the dividing streamline exhibits a strong extra Reynolds number dependence.  To explain this, a toy vorticity diffusion model with an edge layer at the boundary was developed.  The model indicates that the extra Reynolds number dependence arises primarily from role of $\nu$ in determining the thickness of the edge layer.  The dependence on $L/D$ at later times arises mostly from the ring Reynolds number $\Gamma(t)/\nu$ which determines the thickness of the edge layer.  The conventional Gaussian profile, in which $\nu$ enters (mostly) through the product $\nu t$, does not exhibit the extra $\Rey$ dependence.   

\item A diagnostic of the function $\zetah = \zetah(\psih, t)$ was provided by the coefficients of its cubic fit plotted versus $\tstar/\Rey$.  They indicate a lessening with time ($\tstar/\Rey$) of the $L/D$ and (extra) $\Rey$ dependence.  Nevertheless, a non-negligible dependence on both parameters continues to remain, most prominently in the value of $\zetah$ as the dividing streamline is approached, for as long as we have run the simulations.
\end{enumerate}

\section{Suggestions for future work}\label{sec:future}

\begin{enumerate}
\item 
\label{item:destruction} 
An important remaining question is how advective balance is maintained in the face of a diffusion term that acts to destroy it.  Suppose that at some instant we have $\zeta = F(\psi, t)$,  Then upon using the chain rule relations
\ba
   \frac{\p F}{\p x} &=& -u_r r\frac{\p F}{\p \psi}, \hskip0.5truecm \frac{\p F}{\p r} = u_x r\frac{\p F}{\p\psi},\\
   \frac{\p^2 F}{\p x^2} &=& (u_r r)^2\frac{\p^2F}{\p\psi^2} - r\frac{\p u_r}{\p x}\frac{\p F}{\p\psi},\\
   \frac{\p^2 F}{\p r^2} &=& (u_x r)^2\frac{\p^2F}{\p\psi^2} + \frac{\p(u_x r)}{\p r}\frac{\p F}{\p\psi},
\ea
together with
\be
   \zeta = \frac{\omega_\phi}{r} = \frac{1}{r}\left(\frac{\p u_r}{\p x} - \frac{\p u_x}{\p r}\right),
\ee
the vorticity equation \eqp{vort} becomes:
\be
   \frac{\p F}{\p \tau} = r^2q^2\frac{\p^2F}{\p\psi^2} - \left(r^2 F - 4u_x\right)\frac{\p F}{\p\psi}, \eql{axi_diff}
\ee
where $\tau = \nu t$ and $q^2 = u_x^2 + u_r^2$.  The coefficients $r^2 q^2, r^2$, and $u_x$ of the three terms on the right-hand-side of \eqp{axi_diff} are not functions of $\psi$ only.  This should be contrasted with the planar case where streamlines are circular and the diffusion operator is a function of $\psi_{2D}$.  We conclude that the diffusion term in axisymmetric flow acts to destroy advective balance.  How is it then restored?  A reasonable conjecture is that disturbances to the advectively-balanced state induced by the diffusion term are sheared by differential rotation and eventually average out along each closed streamline.  Note that this is an advective effect.  It is analogous to averaging by Taylor (or shear) dispersion of a passive scalar
around closed streamlines studied for planar flow by \cite{Rhines_and_Young_1983}, who conjectured that the same mechanism could occur for vorticity.  When the Reynolds number, $\Gamma(t)/\nu$, becomes sufficiently small, destruction of advective balance by the diffusion term will dominate its restoration by differential rotation.

\item If streamline-averaging accurately describes the restoring effect of advection, then the following algorithm could be used to compute the evolution of $\zeta$: (i) Diffuse $\zeta$ for a short time $\delta t$ imposing the boundary condition dictated by the edge layer.  (ii) To mimic the effect of advection, average $\zeta$ along each closed streamline.  Upon combining the two steps into one, \eqp{axi_diff} becomes
\be
   \frac{\p F}{\p \tau} = \left<r^2q^2\right>\frac{\p^2F}{\p\psi^2} - \left(\left<r^2\right> F - 4\left<u_x\right>\right)\frac{\p F}{\p\psi}, \eql{axi_diff2}
\ee
within the vortex ring bubble.  Here $\left<.\right>$ denotes an average around a closed streamline. (iii) The final step of the algorithm would be to compute the resulting velocity field and streamfunction from the Poisson equation.

\item The most non-trivial aspect of the above procedure is obtaining the boundary condition on $F(\psi)$ where the flow in the interior of the bubble matches with edge layers.
The present work ignored the detailed structure of the edge layer and wake.  They should be studied in the future.  Hints can be obtained from the work of \cite{Harper_and_Moore_1968} on falling drops.  An analytical or semi-analytical treatment that matches the interior flow, edge layers, and exterior potential flow would be most satisfying.
An even more satisfying analysis would also quantify the deviation from advective balance at every location around a closed streamline.  One aspect of this deviation is the forward lean of vorticity contours seen in Figure~\ref{fig:zeta_contours}a (Prof. Brian Cantwell, private communication).  This fore-aft asymmetry may be related to the fact that the windward side of the edge-layer experiences positive strain along the dividing streamline, while the opposite is true on the leeward side.

\item
The Prandtl-Batchelor theorem asserts that $F(\psi) = \mathrm{constant}$ for \textit{exactly steady} axisymmetric flow with closed streamlines \citep[][pp. 186-187]{Batchelor_1956}, a result obtained by integrating the momentum equation around a closed streamline.  The critical assumption is that the rate of change of circulation around every closed streamline is zero.  Even though this is not true in the present case, Batchelor's integral analysis might still be useful.  Note that the recirculation region of a wake behind an axisymmetric bluff body or the interior of a falling drop \citep{Harper_and_Moore_1968} can remain steady, in spite of viscous diffusion, because of a continual source of vorticity at the dividing streamline.  For a bluff body, this source is the separated boundary-layer of the body, and for a drop it is the boundary condition of stress and velocity continuity at the two-fluid interface.  In the present situation, there is a sink of vorticity at the dividing streamline.


\end{enumerate}

\centerline{Acknowledgements}
KS is grateful to Thomas Hartlep and Alan Wray for their suggestions during the internal review.  
We also thank the referees for their useful suggestions.
 
\appendix

\section{Mesh and time-step refinement}\label{sec:refine}

\begin{table}
\begin{center}
\begin{tabular}{l c c}
\toprule
 Grid designation&$N_x \times N_r$&Max. grid spacing\\
                               &                                & in vortex region\\
1x (baseline)         & $1103 \times 251$&$0.01D$\\
2x                            &$2339 \times 531$&$0.005D$\\
\midrule
\end{tabular}
\end{center}
\caption{Mesh refinement parameters.}
\label{tab:refine}
\end{table}
Table~\ref{tab:refine} provides the grid sizes used in the refinement study and Table~\ref{tab:zeta_errors} provides the results.   The 2x grid has twice the resolution of the baseline grid inside the tube and in the vortex region.  
This region is axially and radially shorter than the computational domain in order to keep the number of mesh points manageable.  The quantity compared in Table~\ref{tab:zeta_errors} is $\zetastar$ at the same $100 \times 100$ points used to obtain $\zeta(\psi)$ profiles.  This comparison was performed at three values of $\tstar$ indicated in the tables.
\begin{table}
\begin{tabular}{llllllll}
\multicolumn{2}{c}{}&
\multicolumn{2}{c}{$\tstar = 6$}&
\multicolumn{2}{c}{$\tstar = 12.8$}&
\multicolumn{2}{c}{$\tstar = 18$ or 20}\\
&&Absolute & Relative [\%] & Absolute & Relative [\%] & Absolute & Relative [\%]\\
\hline
\multirow{4}{*}{$L/D = 1$} & Mean &        & 0.44   & & 0.74    &&0.23\\
                                                 &            &        & (0.15) & & (0.93)& & (0.27)\\
                                                & Max.  & 0.030  & 2.71& 0.031      & 1.57& 0.00063 & 0.56\\
                                                &            & (0.018)&(1.31)&(-0.0068)&(1.86)&($-0.0030$)&(0.69)\\
\hline
\multirow{4}{*}{$L/D = 3$} & Mean &       & 0.33&              & 0.071 &       & 0.63\\
                                                 &            &       &(0.82)&            & (0.22)                            &       &(0.16) \\
                                                & Max. & 0.015& 1.19& 0.0042 & 0.23 & $-.0069$ & 1.37\\
                                                 &          &(0.048)&(4.12)&($-0.0023)$&(1.37)&(0.0097)&(0.62)\\
\hline
\end{tabular}

\caption{Change (baseline relative to 2x grid) in $\zetastar(\psi)$ due to mesh and time-step refinement.  Values within parentheses are for time-step refinement ($\Delta \tstar = 0.02$ to $\Delta \tstar = 0.01$) while values outside parenthesis are for mesh refinement.  The columns for relative difference show the mean of the relative absolute difference and the maximum of the relative absolute difference.
$\Rey = 2000$.}
\label{tab:zeta_errors}
\end{table}

Table~\ref{tab:zeta_errors} shows that when a grid refinement is performed, the maximum of the relative change in $\zetastar$ is 2.71\%; it occurs near the dividing streamline where $\zetastar$ is small.  When the time step is halved from $\Delta\tstar = 0.02$ to $\Delta\tstar = 0.01$ this quantity is 4.12\%; this value occurs far from the center of the vortex where the last portion of trailing jet flow is still being entrained and is not representative of errors at later times.


\section{Initial vorticity profile for the one-dimensional model}
\label{sec:ic_model}

For a sharp edge and constant (in time) velocity upstream of the edge, the theory of self-similar roll-up \citep{Kirde_1962} gives the following profile for the circumferential velocity in the vortex at the end of the piston stroke, $t = T$:
\be
   u_\theta(s, T) = C_0 \frac{s}{s_0} M(\threequart, 2; \xi), \hskip 0.5 truecm \xi \equiv -s^2/4\nu T, \hskip0.5truecm 0 \le s \le s_0, \eql{utheta}
\ee
where $s$ is cylindrical radial distance from the core center, $s_0$ is the outermost radius of the vortex spiral at $t = T$, and $M(a, b; x)$ is the confluent hypergeometric function \citep[][p. 503]{Abramowitz_and_Stegun_1965}, and the constant $C_0$ is
\be
   C_0 = u_\theta(s_0) \left[M(\threequart, 2; \xi_0)\right]^{-1}, \displayskip \xi_0 \equiv -s_0^2/4\nu T.
\ee
The diffusion model of the paper requires specification of the vorticity corresponding to \eqp{utheta} which is
\be
   \omega_\phi(s, T) = \frac{1}{s}\frac{\p}{\p s}\left(s u_\theta\right) = \frac{C_0}{s_0} \left[2M(\threequart, 2; \xi) + \frac{3}{4}\xi M(\sevenquart, 3; \xi)\right],
   \hskip0.5truecm 0 \le s \le s_0, 
\ee
with $\omega_\phi = 0$ for $s > s_0$.
Kirde's theory does not consider the profile in the outer region of the core, denoted as region I in \cite{Pullin_1979};
in terminating the velocity profile \eqp{utheta} at $s = s_0$ and setting $\omega_\phi = 0$ for $s > s_0$, we are following equation (2.16) in \cite{Saffman_1978}.  
 

 To specify the radius, $s_0$,  of the vortex spiral and the initial ring radius, $R_0$, we use the
following expressions from \cite{Saffman_1978}:
\be
(s_0)_\mathrm{Saffman} = 0.28 D^{1/3} L^{2/3}, \hskip 0.5truecm R_0 = D/2 + 0.11 D^{1/3} L^{2/3},
\ee
which use scalings from the theory of self-similar roll-up and constants from fitting experiments.
For the 0.28 constant, see the fourth line after Saffman's eq. 3.6, and for the 0.11 constant see the second line on pg. 633 of Saffman's paper.  To prevent the spiral from over-filling the notional dividing streamline, we set
$s_0 = \min\left[(s_0)_\mathrm{Saffman}, s_\rmDS\right]$.

The circumferential velocity at the core boundary is obtained from the circulation, i.e.,
\be
   u_\theta(s_0) = \Gamma_0 / 2\pi s_0,
\ee
where for the circulation $\Gamma_0$ we use
\be
   \frac{\Gamma_0}{\Gamma_\rmslug} = 1.14 + 0.32 (L/D)^{-1}, \eql{Gamma0}
\ee
which is equation (2.6) from \cite{Shariff_and_Leonard_1992} and represents a fit to experiments.  
For the trapezoidal piston velocity prescribed in the present work, the slug circulation is
\be
   \Gamma_\rmslug \equiv \frac{1}{2} \int_0^T U^2_\rmp(t)\, dt =  0.433 U_0^2 T. \eql{Gamma_slug}
\ee
\bibliographystyle{jfm}
\bibliography{paper}

\begin{thebibliography}{32}
\expandafter\ifx\csname natexlab\endcsname\relax\def\natexlab#1{#1}\fi

\bibitem[Abramowitz \& Stegun(1965)]{Abramowitz_and_Stegun_1965}
{\sc Abramowitz, M. \& Stegun, I.A.} 1965 {\em Handbook of Mathematical
  Functions\/}. New York: Dover.

\bibitem[Batchelor(1956)]{Batchelor_1956}
{\sc Batchelor, G.K.} 1956 On steady laminar flow with closed streamlines at
  large {R}eynolds number. {\em J. Fluid Mech.\/} {\bf 1}, 177--190.

\bibitem[Batchelor(1967)]{Batchelor_1967}
{\sc Batchelor, G.K.} 1967 {\em An Introduction to Fluid Dynamics\/}. Cambridge
  University Press.

\bibitem[Berezovski \& Kaplanski(1987)]{Berezovski_and_Kaplanski_1987}
{\sc Berezovski, A.A. \& Kaplanski, F.B.} 1987 Diffusion of a ring vortex. {\em
  Fluid Dyn.\/} {\bf 22}~(6), 832--837, {Transl. of Izvestiya Akademii Nauk,
  Mekhanika Zhidkosti i Gaza}.

\bibitem[Cater {\em et~al.\/}(2004)Cater, Soria \& Lim]{Cater_etal_2004}
{\sc Cater, J.E., Soria, J. \& Lim, T.~T.} 2004 The interaction of the piston
  vortex with a piston-generated vortex ring. {\em J. Fluid Mech.\/} {\bf 499},
  327--–343.

\bibitem[Couder \& Basdevant(1986)]{Couder_and_Basdevant_1986}
{\sc Couder, Y. \& Basdevant, C.} 1986 Experimental and numerical study of
  vortex couples in two-dimensional flows. {\em J. Fluid Mech.\/} {\bf 173},
  225--251.

\bibitem[Eydeland \& Turkington(1988)]{Eydeland_and_Turkington_1988}
{\sc Eydeland, A. \& Turkington, B.} 1988 A computational method of solving
  free-boundary problems in vortex dynamics. {\em J. Comp. Phys.\/} {\bf 78},
  194--214.

\bibitem[Ferziger \& Peri\'c(2002)]{Ferziger_and_Peric_2002}
{\sc Ferziger, J.H. \& Peri\'c, M.} 2002 {\em {C}omputational {M}ethods for
  {F}luid {D}ynamics\/}. Berlin: Springer.

\bibitem[Fukumoto \& Moffatt(2000)]{Fukumoto_and_Moffatt_2000}
{\sc Fukumoto, Y. \& Moffatt, H.~K.} 2000 Motion and expansion of a viscous
  vortex ring. {P}art 1. {A} higher-order asymptotic formula for the velocity.
  {\em J. Fluid Mech.\/} {\bf 417}, 1--45.

\bibitem[Gharib {\em et~al.\/}(1998)Gharib, Rambod \&
  Shariff]{Gharib_etal_1998}
{\sc Gharib, M., Rambod, E. \& Shariff, K.} 1998 A universal time scale for
  vortex ring formation. {\em J. Fluid Mech.\/} {\bf 360}, 121--140.

\bibitem[Harper \& Moore(1968)]{Harper_and_Moore_1968}
{\sc Harper, J.~F. \& Moore, D.~W.} 1968 The motion of a spherical liquid drop
  at high {R}eynolds number. {\em J. Fluid Mech.\/} {\bf 32}~(2), 367--391.

\bibitem[Kambe \& Oshima(1975)]{Kambe_and_Oshima_1975}
{\sc Kambe, T. \& Oshima, Y.} 1975 Generation and decay of viscous vortex
  rings. {\em J. Phys. Soc. of Japan\/} {\bf 38}~(1), 271--280.

\bibitem[Kaplanski {\em et~al.\/}(2012)Kaplanski, Fukumoto \&
  Rudi]{Kaplanski_etal_2012}
{\sc Kaplanski, F., Fukumoto, Y. \& Rudi, Y.} 2012 Reynolds-number effect on
  vortex ring evolution in a viscous fluid. {\em Phys. Fluids\/} {\bf 24}~(3),
  033101:1--13.

\bibitem[Kaplanski \& Rudi(2001)]{Kaplanski_and_Rudi_2001}
{\sc Kaplanski, F.B. \& Rudi, Y.A.} 2001 Evolution of a viscous vortex ring.
  {\em Fluid Dyn.\/} {\bf 36}~(1), 16--25, {Transl. of Izvestiya Akademii Nauk,
  Mekhanika Zhidkosti i Gaza}.

\bibitem[Kaplanski \& Rudi(2005)]{Kaplanski_and_Rudi_2005}
{\sc Kaplanski, F.B. \& Rudi, Y.A.} 2005 A model for the formation of
  ``optimal'' rings taking into account viscosity. {\em Phys. Fluids\/} {\bf
  17}, 087101:1--7.

\bibitem[Kirde(1962)]{Kirde_1962}
{\sc Kirde, K.} 1962 {Untersuchungen \"uber die zeitliche Weiterentwicklung
  eines Wirbels mit Vorgegebener Anfangsverteilung}. {\em Ingenieur-Archiv\/}
  {\bf 31}~(6), 385--404.

\bibitem[Leonard(1979)]{Leonard79}
{\sc Leonard, B.P.} 1979 A stable and accurate convective modelling procedure
  based on quadratic upstream interpolation. {\em Computer Meth. Appl. Mech.
  Engr.\/} {\bf 19}, 59--98.

\bibitem[Linden \& Turner(2001)]{Linden_and_Turner01}
{\sc Linden, P.F. \& Turner, J.S.} 2001 The formation of `optimal' vortex rings
  and the efficiency of propulsive devices. {\em J. Fluid Mech.\/} {\bf 427},
  61--72.

\bibitem[Mohseni \& Gharib(1998)]{Mohseni_and_Gharib98}
{\sc Mohseni, K. \& Gharib, M.} 1998 A model for universal time scale of vortex
  ring formation. {\em Phys. Fluids\/} {\bf 10}, 2436--2438.

\bibitem[Moore(1980)]{Moore_1980}
{\sc Moore, D.~W.} 1980 The velocity of a vortex ring with a thin core of
  elliptical cross section. {\em Proc. Roy. Soc. Lond. A\/} {\bf 370}~(1742),
  407--415.

\bibitem[Nitsche \& Krasny(1994)]{Nitsche_and_Krasny_1994}
{\sc Nitsche, M. \& Krasny, R.} 1994 A numerical study of vortex ring formation
  at the edge of a circular tube. {\em J. Fluid Mech.\/} {\bf 276}, 139--161.

\bibitem[Norbury(1973)]{Norbury_1973}
{\sc Norbury, J.} 1973 A family of steady vortex rings. {\em J. Fluid Mech.\/}
  {\bf 57}, 417--431.

\bibitem[Phillips(1956)]{Phillips_1956}
{\sc Phillips, O.~M.} 1956 The final period of decay of non-homogeneous
  turbulence. {\em Math. Proc. Cambridge Phil. Soc.\/} {\bf 52}~(1),
  135–--151.

\bibitem[Pullin(1979)]{Pullin_1979}
{\sc Pullin, D.} 1979 Vortex ring formation at tube and orifice openings. {\em
  Phys. Fluids\/} {\bf 22}~(3), 401--403.

\bibitem[Rhines \& Young(1983)]{Rhines_and_Young_1983}
{\sc Rhines, P.B. \& Young, W.R.} 1983 How rapidly is a passive scalar mixed
  within closed streamlines? {\em J. Fluid Mech.\/} {\bf 133}, 133--145.

\bibitem[Rott \& Cantwell(1993{\natexlab{{\em a\/}}})]{Rott_and_Cantwell_1993a}
{\sc Rott, N. \& Cantwell, B.} 1993{\natexlab{{\em a\/}}} {Vortex drift. I:
  Dynamic interpretation}. {\em Phys. Fluids A\/} {\bf 5}~(6), 1443--1450.

\bibitem[Rott \& Cantwell(1993{\natexlab{{\em b\/}}})]{Rott_and_Cantwell_1993b}
{\sc Rott, N. \& Cantwell, B.} 1993{\natexlab{{\em b\/}}} {Vortex drift. II:
  The flow potential surrounding a drifting vortical region}. {\em Phys. Fluids
  A\/} {\bf 5}~(6), 1451--1455.

\bibitem[Saffman(1970)]{Saffman_1970}
{\sc Saffman, P.~G.} 1970 The velocity of viscous vortex rings. {\em Stud.
  Appl. Math.\/} {\bf 49}~(4), 371--380.

\bibitem[Saffman(1978)]{Saffman_1978}
{\sc Saffman, P.~G.} 1978 The number of waves on unstable vortex rings. {\em J.
  Fluid Mech.\/} {\bf 84}~(4), 625--–639.

\bibitem[Shariff \& Leonard(1992)]{Shariff_and_Leonard_1992}
{\sc Shariff, K. \& Leonard, A.} 1992 Vortex rings. {\em Ann. Rev. Fluid
  Mech.\/} {\bf 24}, 235--279.

\bibitem[Shusser \& Gharib(2000)]{Shusser_and_Gharib00}
{\sc Shusser, M. \& Gharib, M.} 2000 Energy and velocity of a forming vortex
  ring. {\em Phys. Fluids\/} {\bf 12}~(3), 618--621.

\bibitem[Stanaway {\em et~al.\/}(1988)Stanaway, Cantwell \&
  Spalart]{Stanaway_etal_1988a}
{\sc Stanaway, S., Cantwell, B.J. \& Spalart, P.R.} 1988 A numerical study of
  viscous vortex rings using a spectral method. Technical Memorandum 101041.
  NASA.

\end{thebibliography}
\end{document}